  \documentclass[twocolumn]{aastex631}
\usepackage{amsmath}
\usepackage{tabularx}
\usepackage{graphicx}
\usepackage{booktabs}       
\usepackage{amsfonts}       
\usepackage{nicefrac}       
\usepackage{microtype}
\usepackage{cleveref}
\usepackage{multirow}
\usepackage{makecell}
\usepackage{array}
\usepackage[caption=false]{subfig}
\usepackage{stfloats}
\usepackage[utf8]{inputenc}

\renewcommand{\arraystretch}{1.2}
\shorttitle{Magnetic Hybrid stars}
\shortauthors{I. A. Rather et al.}
\graphicspath{{./}{Paper_Plots/}}

\begin{document}

\title{Magnetic-Field Induced Deformation in Hybrid Stars}

\author[0000-0001-5930-7179]{Ishfaq A. Rather}
\email{ishfaq.rather@tecnico.ulisboa.pt}
\affiliation{Centro de Astrof{\'i}sica e Gravita{\c c}{\~a}o-CENTRA, \\
Instituto Superior T{\'e}cnico-IST, 
Universidade de Lisboa-UL, \\
	Av.~Rovisco Pais, 1049-001 Lisboa, Portugal}
\author[0000-0003-1126-7731]{Asloob A. Rather}
\affiliation{Department of Physics, \\
Aligarh Muslim University, Aligarh 202002, India}
\affiliation{Department of the School of Education,\\
Government of the Union Territory of Jammu and Kashmir, India}
\author[0000-0002-5011-9195]{Il{\'i}dio Lopes}
\affiliation{Centro de Astrof{\'i}sica e Gravita{\c c}{\~a}o-CENTRA, \\
Instituto Superior T{\'e}cnico-IST, 
Universidade de Lisboa-UL, \\
	Av.~Rovisco Pais, 1049-001 Lisboa, Portugal}

\author[0000-0001-5578-2626]{V. Dexheimer}
\affiliation{Department of Physics, \\
	Kent State University, Kent, OH 44242, USA}

\author[0000-0002-5785-6526]{A. A. Usmani}
\affiliation{Department of Physics, \\
Aligarh Muslim University, Aligarh 202002, India}

\author[0000-0003-4616-5794]{S. K. Patra}
\affiliation{Institute of Physics, Bhubaneswar 751005, India}
\affiliation{Homi Bhabha National Institute, \\
	Training School Complex, Anushakti Nagar,\\
	 Mumbai 400094, India}



\begin{abstract}
The effects of strong magnetic fields on the deconfinement phase transition expected to take place in the interior of massive neutron stars are studied in detail for the first time. For hadronic matter, the very general density-dependent relativistic mean-field (DD-RMF) model is employed, while the simple, but effective vector-enhanced bag model (vBag) model is used to study quark matter. Magnetic-field effects are incorporated into the matter equation of state and in the general-relativity solutions, which also satisfy Maxwell's equations. We find that for large values of magnetic dipole moment, the maximum mass, canonical mass radius, and dimensionless tidal deformability obtained for stars using spherically symmetric Tolman–Oppenheimer–Volkoff (TOV) equations and axisymmetric solutions attained through the LORENE library differ considerably. The deviations depend on the stiffness of the equation of state and on the star mass being analyzed. This points to the fact that, unlike what was assumed previously in the literature, magnetic field thresholds for the approximation of isotropic stars and the acceptable use of TOV equations depend on the matter composition and interactions.

\end{abstract}



\section{Introduction} \label{sec:intro}

Matter under extreme densities, temperatures, and magnetic fields is among the most popular and fascinating current topics of research. Neutron stars (NSs) provide the perfect environment to study physics under extreme conditions. Recent advances in NS observation have yielded intriguing conclusions regarding their maximum mass, canonical mass ($1.4$ $M_{\sun}$), radius, tidal deformability, and so on  \citep{PhysRevLett.119.161101,PhysRevLett.121.161101,Abbott_2021,Miller_2019a,Riley_2019,Miller_2021,Riley:2021pdl}. However, there is still a large uncertainty concerning the equation of state (EoS) of dense matter. The density inside NSs reaches several times nuclear saturation density ($\rho_{0}=10^{14}$ g cm$^{-3}=0.15$ fm$^{-3}$), making it difficult to determine their inner structure. As interior densities approach $10^{15}$ g cm$^{-3}$, theoretical models describing cold and dense matter, calibrated around $\rho_{0}$ for isospin-symmetric nuclear matter (SNM), must be extrapolated in both density and isospin asymmetry. To test these ideas, the structure of NSs is computed starting from the energy–momentum tensor, solving equations for hydrostatic equilibrium, and comparing them to astrophysical observations. One constraint that all NS models must meet is the ability to explain the highest measured NS mass, which according to recent astrophysical data is above $2$ $M_{\sun}$ \citep{Antoniadis1233232,2021ApJ...915L..12F}. \par

The existence of these massive NSs rules out a number of soft EoSs, in which the expected maximum mass of the compact object is less than the observed maximum mass. Meanwhile, it has been long accepted in the nuclear physics community that hyperons emerge in dense medium under NS conditions at a few times $\rho_{0}$ \citep{1960SvA.....4..187A,GLENDENNING1982392,1985ApJ...293..470G}.
 The EoS softens in the presence of hyperons ruling out the occurrence of $2$ $M_{\sun}$ NSs in many models. To prevent this, the EoS without hyperons can be very stiff but that would lead to a large stellar radii. 
Together, all of this gives rise to the so-called hyperon puzzle, which is a tremendous challenge in nuclear physics  \citep{Chamel:2013efa}. A possible way to solve this problem would be to consider deconfinement to quark matter, as discussed in the following. \par

NSs are not just extremely dense objects, but they are also known to present extremely strong magnetic fields. 
But a few X-ray isolated NSs and rotating radio transients present even stronger magnetic fields. Soft gamma repeaters (SGRs) and anomalous X-ray pulsars (AXPs) present the strongest magnetic fields found in NSs with surface values of the order 10$^{14}$-10$^{15}$ G \citep{2015SSRv..191..315M,doi:10.1146/annurev-astro-081915-023329,Harding_2006,Turolla_2015}. 
But, more interestingly, data from the source 4U 0142+61 for slow phase modulations
in hard X-ray pulsations (interpreted as free precession)
suggests magnetic fields of the order of $10^{16}$ G~\citep{Makishima:2014dua,DallOsso:2018dos} inside this pulsar. Magnetic fields of the order of $10^{16}$ G or higher are also expected to be produced in NS mergers \citep{Ciolfi:2013dta,Giacomazzo:2014qba,Ciolfi2020,Palenzuela:2021gdo,Ruiz:2021qmm}.

Because the maximal magnetic field in the interior of magnetars cannot be measured directly, it is commonly predicted using the virial theorem - most estimates lead to a theoretical maximum of the order of $10^{18}$ G \citep{1991ApJ...383..745L}. A strong magnetic field can have several implications for NSs, such as modifying the EoS due to Landau quantization of the constituent charged particles \citep{zbMATH03285978,PhysRevD.86.125032},
changing the energy–momentum tensor, and breaking the stellar spherical symmetry.
Several studies have shown the effect of the magnetic field on the NS EoS and on stellar properties \citep{PhysRevLett.78.2898,Broderick_2000,Rabhi_2008,PhysRevC.89.045805,PhysRevLett.79.2176,Gomes:2017zkc,Pili:2017yxd,Felipe:2007vb,PhysRevC.89.015805,Dexheimer2021,delta_baryon_ns,Gomes19,PhysRevC.94.055805,10.1093/mnras/stu2677,Tolos_2016,PhysRevC.99.065803,latent_menz,PhysRevC.99.055811,10.1093/mnras/stu2706}. Most of the previous studies that include  magnetic field effects on the EoS use isotropic Tolman–Oppenheimer–Volkoff (TOV) equations to determine the relevant stellar properties. However, it should be emphasized that for strong magnetic fields, below the threshold at which Landau quantization effects on the EoS become non-negligible, the deviations from spherical symmetry can already be  considerable \citep{Gomes19,10.1093/mnras/stu2706}. Strong magnetic fields can substantially deviate NS configurations from spherical symmetry, and in this case, spherically symmetric TOV equations can no longer be used to describe their macroscopic structure. \par

Concerning the EoS, in a density regime, which ab initio methods fail to describe, relativistic mean field (RMF) models have been successful in characterizing both finite and infinite nuclear matter \citep{Walecka:1974qa}. The primary mechanism in this case consists of hadrons interacting through a mean field of mesons. The use of different mesons such as $\sigma$, $\omega$, $\rho$, and $\delta$ has improved the predictions of SNM properties and constrained them well inside the expected regime \citep{BOGUTA1977413,SEROT1979146,SUGAHARA1994557,PhysRevLett.86.5647}. The density-dependent RMF (DD-RMF) model replaces the self- and cross-coupling of various mesons in the classic RMF model with density-dependent coupling constants \citep{PhysRevLett.68.3408}. Parameter sets like DD-ME1 \citep{PhysRevC.66.024306} and DD-ME2 \citep{PhysRevC.71.024312} generate very massive NSs with a $2.3-2.5$ $M_{\sun}$ maximum mass. Several new DD-RMF parameter sets such as DD-LZ1 \citep{ddmex} and DD-MEX \citep{TANINAH2020135065} also produce stiff EoSs. Note that the density-dependent functional DD2 EoS \citep{Typel:2009sy,TYPEL1999331,Banik:2014qja} with excluded volume correction \citep{Typel:2016srf}, which accounts for the finite size of nucleons caused by the repulsive attraction between their internal quarks caused by Pauli blocking effects, is widely used in simulations of, for example, binary NS mergers \citep{PhysRevD.93.124046,Radice_2018,Lehner_2016,Tootle:2022pvd} and supernovae \citep{Fischer:2021tvv,Jakobus:2022ucs}. 

The idea of deconfined quark matter in the interior of NSs has been studied since the 1980s \citep{PhysRevD.30.272}. An NS with hadronic matter in the core followed by a phase transition to quark matter at least a couple of times saturation density is referred to as a hybrid star. To study quark matter in NS cores, several different models have been used. The simple MIT bag model \citep{PhysRevD.9.3471,PhysRevD.17.1109,PhysRevD.30.2379} was the first realistic description proposed to study pure quark and hybrid stars. The  Nambu-Jona-Lasinio (NJL) model, which originally described the mass gap in the Dirac spectrum of nucleons (in analogy to the BCS mechanism for superconductivity) \citep{PhysRev.122.345,PhysRev.124.246}, was modified to describe interacting quark matter \citep{Kleinert:1976ds,Volkov:1984kq,Hatsuda:1984jm}. Somehow between these, the modified version of the Bag model, vector-enhanced bag model (vBag) \citep{Kl_hn_2015}, was introduced as an effective model for studying astrophysical processes. This model is preferred over the simple bag model because it takes into consideration repulsive vector interactions as well as dynamic chiral symmetry breaking (D$\chi$SB). The repulsive vector interaction and the introduction of a mixed phase allow one to reproduce hybrid stars with masses larger than the $2$ $M_{\sun}$ limit.\par

An alternative for a sharp first-order phase transition emerges when one considers that the surface tension of quark matter is not infinite. This so-called non-congruent phase transition gives rise to a mixture of phases in which quantities such as electric charge neutrality are fulfilled globally \citep{Glendenning:1992vb}. In astrophysics, this is referred to as a Gibbs construction and generates a phase in which hadrons \textit{melt} into their constituent quarks gradually over kilometers inside NSs. See Ref.~\citep{Hempel:2013tfa} and references therein for an extended review of the topic. A more thorough treatment of the mixed phase, including contributions from surface and Coulomb effects requires accurate knowledge of the surface tension between two phases, which is still unknown, having an estimated value ranging from 10-100 MeV fm$^{-2}$ \citep{PhysRevD.30.2379,Kajantie:1990bu,Alford:2002rj,Lugones:2013ema,Lugones:2016ytl}. The Gibbs construction produces results that are somehow similar to those obtained with the lower value of the abovementioned surface tension range, and so the contribution from surface and Coulomb effects can be ignored (when calculating properties such as stellar masses and radii) \citep{PhysRevD.76.123015,VOSKRESENSKY200293}. In this work, we study hybrid stars by means of a Gibbs construction that gives rise to a mixed phase connecting the hadronic phase described by the DD-RMF model and the quark phase described by the vBag model.

When the magnetic field effects in hadronic and quark phases are compared, it is found that changes in the population of charged particles in the hadronic phase are larger than for uncharged particles, noting that all quarks have an electric charge \citep{PhysRevC.89.015805,Rabhi_2009,2014JPhG...41a5203D}. Also, a difference due to particle masses is observed with baryons having a substantially larger mass than quarks and leptons \citep{Peterson:2021teb}. These two characteristics imply that magnetic field effects should be more pronounced in quark matter. On the other hand, it has been shown that changes in the amount of (baryonic) charge fraction affect more strongly the hadronic phase than the quark phase \citep{Aryal:2020ocm}

To understand better how these effects affect macroscopic stellar properties, we use our EoS with quark deconfinement in the publicly available Language Objet pour la RElativit\'e Num\'eriquE (LORENE) library \citep{LORENE,10.1093/mnras/stu2706} for the first time to investigate in detail quark deconfinement. LORENE solves the coupled Einstein-Maxwell field equations in order to determine stable anisotropic magnetic-star configurations. However, to highlight and calculate the deviation from spherical symmetry produced in stellar properties, we also analyze results from TOV solutions.

Our paper is organized as follows: 
 the hadronic and quark matter EoSs, the mixed phase region, and the deconfinement phase transition properties together with the pure electromagnetic contribution to the energy-momentum tensor, are discussed in Sec.~\ref{sec1}. NS structure, which contains the TOV equations along with a description of the LORENE library, and a discussion of the magnetic field distribution are discussed in Sec.~\ref{sec3}. The results of magnetic field effects on the EoS with quark deconfinement and hybrid stars are explained and discussed in Sec.~\ref{sec6}. Finally, a summary and concluding remarks are presented in Sec.~\ref{summary}. 

\section{Microscopic Formalism}
\label{sec1}

\subsection{Hadronic matter}

The most basic and simple RMF Lagrangian involves the scalar-isoscalar $\sigma$ and vector-isoscalar $\omega$ mesons without higher-order interactions \citep{walecka}, resulting in a huge SNM incompressibility $K_0$ \citep{Walecka:1974qa}. A nonlinear self-coupling of the $\sigma$ field was then incorporated by Boguta and Bodmer \citep{BOGUTA1977413}, which reduced the value of SNM incompressibility to realistic levels. The contribution from the vector-isovector $\rho$ and scalar-isovector $\delta$ mesons was later introduced to study isospin asymmetry effects.

The various nonlinear meson coupling terms present in the RMF models are replaced by the density-dependent nucleon-meson coupling constants in the DD-RMF \citep{PhysRevLett.68.3408,PhysRevC.66.024306,PhysRevC.71.024312,ddmex,TANINAH2020135065,Typel:2009sy}, Density-Dependent Relativistic Hartree \citep{PhysRevLett.68.3408} and also in the Density-Dependent Relativistic Hartree-Fock (DD-RHF) \citep{PhysRevC.36.380,PhysRevC.18.1510,LONG2006150} models. Their couplings allow for a consistent study of NSs and yield results that are comparable to other more complicated models. The contribution of the rearrangement term self energies to DD-RMF field equations is the most significant difference from normal RMF models and ensures thermodynamical consistency.

The hadronic matter DD-RMF matter Lagrangian density (with a free Fermi gas of leptons) is given by
\begin{align}\label{lag}
\mathcal{L}_m & =\sum_{b} \bar{\psi}_{b} \Biggl\{\gamma_{\mu}\Bigg(iD^{\mu}-g_{\omega}(\rho_b)\omega_{\mu}-\frac{1}{2}g_{\rho}(\rho_b)\rho_{\mu}\tau\Bigg) \nonumber \\
&-\Bigg(M_b-g_{\sigma}(\rho_b)\sigma\Bigg)\Biggr\} \psi_{b} 
+\frac{1}{2}\Bigg(\partial^{\mu}\sigma \partial_{\mu}\sigma-m_{\sigma}^2 \sigma^2\Bigg) \nonumber \\
&-\frac{1}{4}W^{\mu \nu}W_{\mu \nu}
+\frac{1}{2}m_{\omega}^2 \omega_{\mu} \omega^{\mu}
-\frac{1}{4}R^{\mu \nu} R_{\mu \nu} 
+\frac{1}{2}m_{\rho}^2 \rho_{\mu} \rho^{\mu} \nonumber \\
&+ \sum_l \bar{\psi}_l (i\gamma_{\mu} D^{\mu}-m_l)\psi_l\ ,
\end{align}
where $b$ sums over the baryon octet ($n,p,\Lambda, \Sigma^+, \Sigma^0, \Sigma^-, \Xi^0, \Xi^- $)  and $l$ over leptons $e^-$ and $\mu^-$. $\psi_b$ and  $\psi_l$ represent the baryon and lepton Dirac fields and $M_b$ and  $m_l$ the baryon and lepton masses, respectively. $\tau$ and $\gamma_{\mu}$ denote the baryon isopsin projection operator and the 4-dimensional Dirac matrices, respectively. 

The mesonic tensor fields and covariant derivatives are defined as 
\begin{align}
W^{\mu \nu}&=\partial^{\mu}W^{\nu}-\partial^{\nu}W^{\mu},\nonumber \\
R^{\mu \nu}&=\partial^{\mu}R^{\nu}-\partial^{\nu}R^{\mu}, \nonumber \\
D^{\mu} &= \partial^{\mu}+iQA^{\mu},
\end{align}
where $W_{\mu \nu}$ and $R_{\mu \nu}$ are the antisymmetric tensor fields of $\omega$ and $\rho$ vector mesons. $\sigma$ is a scalar meson. $A^{\mu}$ is the photon field that couples to baryons and leptons with electric charge $Q$.

The isoscalar density-dependent coupling constants for the DD-RMF parameter set are written as a function of baryon (number) density $\rho_b$
\begin{equation}
g_i(\rho_b) = g_i(\rho_0) f_i(x)\ ,
\end{equation}
where the function $f_i(x)$ is given by
\begin{equation}\label{eq5}
f_i(x) = a_i \frac{1+b_i (x+d_i)^2}{1+c_i(x+d_i)^2},\ i=\sigma,\omega\ ,
\end{equation}
with $ x=\rho_b/\rho_{0}$. Additional constraints for the function $f_i(1)=1$,$f^{''}_{\sigma}(1)=f^{''}_{\omega}(1)$, $f^{''}_i(0)=0$ reduce the number of free parameters from eight to three in the Eq.~\ref{eq5}. Among them, the first two constraints are
\begin{equation}
a_i=\frac{1+c_i(1+d_i)^2}{1+b_i(1+d_i)^2},\ 
3c_id_i^2=1\ .
\end{equation}
For the isovector $\rho$ and $\delta$ mesons, the density-dependent coupling constants are given by an exponential dependence
\begin{equation}
g_i(\rho_b)=g_i(\rho_0)\exp[-a_i(x-1)]\ .
\end{equation}

The coupling constants of the nucleons to the $\sigma$ and $\omega$ mesons at saturation are determined from fitting saturation density and binding energy for SNM. The coupling constant of the nucleons to the $\rho$ meson at saturation is fitted by reproducing the empirical saturation properties of nuclear matter, such as the symmetry energy. 
The coupling constants of the hyperons to the vector mesons at saturation are determined from SU(6) symmetry
\begin{align}\label{eq20}
\frac{1}{2}g_{\omega\Lambda}&=\frac{1}{2}g_{\omega\Sigma}=g_{\omega\Xi}=\frac{1}{3}g_{\omega N},\nonumber \\
\frac{1}{2}g_{\rho\Sigma}&=g_{\rho\Xi}=g_{\rho N},\ \  g_{\rho \Lambda}=0\ . 
\end{align}
The coupling constants of the hyperons to the $\sigma$ meson at saturation are determined by fitting the $\Lambda$ hyperon optical potential for SNM
\begin{equation}
U_{\Lambda}^N(\rho)=g_{\omega \Lambda}\omega_0 +\sum_{R}-g_{\sigma \Lambda}\sigma_0\ , 
\end{equation}
to results obtained from lattice calculations~\citep{Inoue:2019jme,Inoue:2018axd}, 
reproducing the following potentials: 
$U_{\Lambda}^N(\rho_0)=-30$ MeV,
$U_{\Sigma}^N(\rho_0)=+30$ MeV,
and $U_{\Xi}^N(\rho_0)=-14$ MeV.
They correspond to the following coupling values at saturation: $g_{\sigma \Lambda}/g_{\sigma N} = 0.6105$, $g_{\sigma \Xi}/g_{\sigma N} = 0.3024$, and $g_{\sigma \Sigma}/g_{\sigma N} = 0.4426$.

For a uniform magnetic field locally pointing in the $z$-direction, $B=B\hat{z}$, the Fermi energy of a charged baryons and leptons, respectively, becomes \citep{Broderick_2000}
\begin{equation}
E_{b}^*=\sqrt{k_{zb}^2 +M_b^{*2}+2\nu|Q_b|B}\ ,
\end{equation}
\begin{equation}
E_{l}=\sqrt{k_{zl}^2 +m_l^{2}+2\nu|Q_l|B}\ ,
\end{equation}
where the baryon effective mass is $M_b^*=M_b-g_\sigma(\rho_b)\sigma$. The quantity $\nu=\Big(n+\frac{1}{2}-\frac{1}{2}\frac{q}{|q|} \sigma_z\Big)=0,1,2,...$ indicates the Landau levels, $n$ is the orbital angular momentum quantum number, and $\sigma_z$ the Pauli matrix.
The highest value of $\nu$ is obtained under the condition that the Fermi momentum of each particle is real. This gives us
\begin{align}
\nu_{max}&=\Bigg\lfloor\frac{E_{b}^{*2} -M_{b}^{*2}}{2|Q_{b}|B}\Bigg\rfloor, \nonumber \\
\nu_{max}&=\Bigg\lfloor\frac{E_l^2 -m_{l}^{2}}{2|Q_{l}|B}\Bigg\rfloor\ , 
\label{landau}
\end{align}
for charged baryons and leptons, respectively.\par

With all baryons from the octet and leptons included, the NS chemical-equilibrium conditions between different particles are
\begin{align}\label{eq13}
&\mu_b=\mu_n =\mu_{\Sigma^0} =  \mu_{\Xi^0}, \nonumber \\
&\mu_p =\mu_{\Sigma^+}=\mu_n -\mu_e, \nonumber \\
&\mu_{\Sigma^-} =\mu_{\Xi^-}=\mu_n +\mu_e,\nonumber \\
&\mu_{\mu} =\mu_e\ .
\end{align}
The (electric) charge neutrality condition is as follows:
\begin{equation}
\rho_p +\rho_{\Sigma^+}= \rho_e+\rho_{\mu^-}+\rho_{\Sigma^-}+\rho_{\Xi^-}\ .
\end{equation}

The EoS and equations of motion in the presence of a magnetic field are discussed in the Appendix (\ref{A}).


\subsection{Quark matter}

The MIT bag model has been extensively used to describe quark matter in NSs. In the original description \citep{PhysRevD.30.2379,PhysRevD.9.3471,PhysRevD.17.1109}, quarks are considered to be free inside a bag and thermodynamic properties are simply derived from a free Fermi gas model. The vBag model \citep{Kl_hn_2015}, a modified version of the bag model, was introduced as an effective model for studying astrophysical processes. It is preferred over the simple bag model because it takes into consideration D$\chi$SB as well as repulsive vector interactions. The repulsive vector interaction is important as it permits hybrid stars to achieve the $2$ $M_{\sun}$ maximum mass limit \citep{vbageos,Rather_2021,Rather_2021a,Lopes:2020btp,Dexheimer:2020rlp,Kumar:2022byc} and hence, satisfy the $2$ $M_{\sun}$ mass constraints \citep{Antoniadis1233232,2021ApJ...915L..12F}.

The Lagrangian density for the vBag model (with a free Fermi gas of leptons) reads as
\begin{align}
\mathcal{L} &=\sum_f  [ \psi_f (i\gamma_{\mu} \partial_{\mu} -m_f -B_{bag}) \psi_f] \Theta_H  \nonumber \\
& -G_V \sum_f (\bar{\psi_f} \gamma_{\mu} \psi_f)^2
+ \sum_l  \psi_l \gamma_{\mu} (i \partial_{\mu} -m_l )\psi_l \ ,
\end{align}
where $u$, $d$, and $s$ quarks and $e^-$ and $\mu^-$ leptons are denoted by subscripts $f$ and $l$, respectively. $B_{bag}$ denotes the bag constant and $\Theta_H$ is the Heaviside step function which allows for the confinement/deconfinement of the bag \citep{PhysRevD.30.2379}. The vector interaction is introduced via the coupling of vector-isoscalar meson to the quarks with coupling constant $G_V$. Quarks $u$, $d$, and $s$ with mass $m_u$ = $m_d$ = 5 and $m_s$ = 100 MeV are considered.

The total energy density and pressure are
\begin{equation}
\mathcal{E}_Q = \sum_{l} \mathcal{E}_l+\sum_{f=u,d,s} \mathcal{E}_{\rm{vBag},f}-B_{dec}\ ,
\end{equation}
\begin{equation}
P_Q = \sum_{l} \mathcal{E}_l+\sum_{f=u,d,s} P_{\rm{vBag},f}+B_{dec}\ ,
\end{equation}
where $B_{dec}$ represents the deconfined bag constant, which lowers the energy per particle, and thus, favors stable strange matter. The energy density and pressure of a single quark flavor are defined as
\begin{equation}\label{q1}
\mathcal{E}_{\rm{vBag},f} = \mathcal{E}_f(\mu_f^*)+\frac{1}{2}K_{\nu}n_f^2 (\mu_f^*)+B_{\chi,f}\ ,
\end{equation}   
\begin{equation}
P_{\rm{vBag},f}= P_f(\mu_f^*)+\frac{1}{2}K_{\nu}n_f^2 (\mu_f^*)-B_{\chi,f}\ .
\end{equation}
The expressions for the free quark matter energy density and pressure in the presence of a magnetic field are shown in the Appendix \ref{A}. The coupling constant parameter $K_{\nu}$ results from the vector interactions and controls the stiffness of matter~\citep{Wei_2019}. In the present study, the coupling constant parameter $K_{\nu}$ is fixed at $4$ GeV$^{-2}$ for a three flavor configuration. The bag constant for a single quark flavor is denoted by $B_{\chi,f}$.

The effective chemical potential $\mu_f^*$ of the system and the quark density are defined as
\begin{equation}
\mu_f^* =\mu_f -K_{\nu}n_f(\mu_F^*)\ .
\end{equation}  
\begin{equation}
n_f (\mu_f) = n_f (\mu^*)
\end{equation}
In order for the phase transition to occur at the same chemical potential $\mu_B=\mu_u+2\mu_d$ for all flavors, the effective bag constant $B_{\rm{eff}}$ is defined in the vBag model as 
\begin{equation}
B_{\rm{eff}}=\sum_{f=u,d,s}B_{\chi,f}-B_{dec}\ .
\end{equation}
Two different values of effective bag constant are used in this work $B_{eff}^{1/4}=130$ and $160$ MeV, and in the final discussion, $180$ MeV is also considered.\par

In the presence of a magnetic field, the transverse component of the momentum of all quarks is quantized into Landau levels
\begin{equation}
E_f = \sqrt{k_{z,f}^2 + m_f^2 + 2\nu|Q_f|B }\ ,
\end{equation}
For quark matter, we obtain the highest value of Landau levels, similar to Eq.~\ref{landau} as
\begin{equation}
\nu_{max} = \Bigg\lfloor\frac{E_f^2 -m_{f}^{2}}{2|Q_{f}|B}\Bigg\rfloor\ , 
\end{equation}

The charge neutrality and chemical-equilibrium conditions for the quark matter are
\begin{equation}
\frac{2}{3}\rho_u -\frac{1}{2}(\rho_d+\rho_s)-\rho_e-\rho_\mu =0,
\end{equation}
\begin{align}
&\mu_s=\mu_d=\mu_u+\mu_e \\
&\mu_\mu=\mu_e.
\end{align}


\subsection{Mixed Phase}
\label{sec3}
In this work we assume that the deconfinement phase transition is of first order and that the quark-matter surface tension is low enough for a mixture of phases to appear. By means of a Gibbs construction \citep{Glendenning:1992vb}, charge neutrality is achieved globally, while the hadronic matter is positively charged and quark matter is negatively charged. Within chemical equilibrium, the stiffness of the phase determines the extension of the mixed phase in density.\par

The expressions for the chemical potential and pressure within the mixed phase are defined as
\begin{equation}\label{e1}
\mu_{b,H} = \mu_{b,Q};\ \  \mu_{e,H} = \mu_{e,Q}\ ,
\end{equation}
and
\begin{equation}\label{e2}
P_{H}(\mu_b,\mu_e) = P_{Q}(\mu_b,\mu_e) = P_{MP}\ ,
\end{equation} 
where the subscripts $H$, $Q$, and $MP$ represent the hadronic phase, quark phase, and the mixed phase, respectively, all containing lepton contributions. From the global charge conservation of the electric and baryon charges, we have
\begin{align}
&\chi \rho_{Q}+(1-\chi) \rho_{H}, \nonumber \\
&\chi \rho_{b_{Q}} +(1-\chi)\rho_{b_{H}}=\rho_{b_{MP}}\ .
\end{align}  
Here, $\chi$ is the quark phase volume fraction given by $\chi= V_Q/(V_Q +V_H)$ and $(1-\chi)$ is the hadronic phase volume fraction. The charge densities in the hadronic and quark phases are represented by $\rho_Q$ and $\rho_H$.

The total energy density of the mixed phase then reads as
\begin{equation}\label{e3}
\mathcal{E}_{MP} = \chi \mathcal{E}_{Q} +(1-\chi)\mathcal{E}_{H}\ ,
\end{equation}
The mixed phase is characterized by a value of $\chi$ which varies from 0, the onset of mixed phase to 1, the onset of the pure quark phase. The equations above determine the properties of the mixed phase, and combined with the equations for hadronic and quark phases, allow us to calculate macroscopic properties for hybrid stars.\par

\subsection{Electromagnetic Contribution}
\label{sub4}
For the pure electromagnetic part, the Lagrangian density is written as
\begin{equation}
\mathcal{L}_\gamma=-\frac{1}{16\pi} F_{\mu \nu}F^{\mu \nu}\ ,
\end{equation}
where $F_{\mu \nu}$ is the electromagnetic field tensor, $F_{\mu \nu}$=$\partial_{\mu}A_{\nu}-\partial_{\nu}A_{\mu}$. Hence, the total Lagrangian density in the presence of a magnetic field is
\begin{equation}
\mathcal{L}=\mathcal{L}_m +\mathcal{L}_\gamma\ ,
\end{equation}
where ``m'' stands for matter, including the hadronic, quark, and mixed phases.

For the energy density and pressure in the presence of a magnetic field $B$, the expressions can be obtained by solving the energy-momentum tensor relation
\begin{equation}
T^{\mu\nu} = T_m^{\mu\nu}+T_\gamma^{\mu\nu}, 
\end{equation}
where \citep{PhysRevD.81.045015,PhysRevD.65.056001}
\begin{align}\label{eq11}
T_m^{\mu\nu} &= \mathcal{E}_m u^{\mu}u^{\nu}-P(g^{\mu\nu}-u^{\mu}u^{\nu}) \nonumber \\
&+\mathcal{M}B \Bigg(g^{\mu\nu}-u^{\mu}u^{\nu}+\frac{B^{\mu}B^{\nu}}{B^2}\Bigg), \nonumber \\
T_\gamma^{\mu\nu} &= \frac{B^2}{4\pi}\Bigg(u^{\mu}u^{\nu}-\frac{1}{2}g^{\mu\nu}\Bigg)-\frac{B^{\mu}B^{\nu}}{4\pi}\ .
\end{align}
Here, $\mathcal{M}$ is the magnetization per unit volume and $B^{\mu}B_{\mu}=-B^2$ where $B^{\mu}$ = $\epsilon^{\mu \nu \alpha \beta}F_{\nu \alpha} u_{\beta}/2$ with $\epsilon^{\mu \nu \alpha \beta}$ being the totally antisymmetric Levi-Civita tensor, and $B$ = $\mathrm{|\textbf{B}|}$. The signature of the metric tensor is $g^{\mu \nu}$=diag(1, -1, -1, -1). The field contribution to the energy-momentum tensor takes the form $T_\gamma^{\mu\nu}$ = diag($B^2/2$, $B^2/2$, $B^2/2$, -$B^2/2$). For matter in  the presence of a magnetic field, the single particle energies of all charged baryons, quarks, and leptons are quantized in the direction perpendicular to the magnetic field. 

The expressions for the matter energy density, pressure, and baryon density obtained in the presence of a magnetic field are shown in Appendix \ref{A}. From Eq. \ref{eq11}, the total energy density is
\begin{equation}
\mathcal{E}=\mathcal{E}_m +\frac{B^2}{8\pi}\ .
\end{equation}
The total pressure in the perpendicular and the parallel directions to the local magnetic field are
\begin{align}
P_{\perp}&=P_m-\mathcal{M}B +\frac{B^2}{8\pi},\nonumber \\
P_{\parallel}&=P_m-\frac{B^2}{8\pi}\ ,
\end{align}
where the magnetization is calculated as
\begin{equation}
\mathcal{M}=\partial P_m/\partial B\ .
\end{equation}

When we use TOV solutions for macroscopic stellar properties, we only consider the pressure in the perpendicular direction to determine the maximum possible NS mass obtained with the magnetic field. When we use solutions from LORENE, we do not account for magnetic field effects in the EOS, as this was shown not to modify significantly stellar masses and radii \citep{10.1093/mnras/stu2706,Franzon:2015sya}.


\section{NS Structure}
\label{sec3}
\subsection{TOV}

The properties of spherically static NSs are obtained using the well-known TOV coupled differential equations given by~\citep{PhysRev.55.364,PhysRev.55.374}
\begin{equation}\label{tov1}
\frac{dP(r)}{dr}= -\frac{[\mathcal{E}(r) +P(r)][M(r)+4\pi r^3 P(r)]}{r^2\bigg(1-\frac{2M(r)}{r}\bigg) }\ ,
\end{equation}
and
\begin{equation}\label{tov2}
\frac{dM(r)}{dr}= 4\pi r^2 \mathcal{E}(r)\ ,
\end{equation}
where $M(r)$ represents the gravitational mass inside radius $r$. The boundary conditions  $P(0)=P_c$ and $M(0)=0$ at the center and $P(R)=0$ at the surface allow one to solve the differential equations and determine the properties of an NS for each given central pressure $P_c$. \par

\subsection{LORENE}

For strong magnetic fields, the spherically symmetric solutions obtained by solving the TOV equations lead to an overestimation of the mass and underestimation of the equatorial radius (when the pressure in the local perpendicular direction to the magnetic field is applied in all directions), and hence, cannot be used for determining stellar properties. For this reason, we use the LORENE library \citep{LORENE,10.1093/mnras/stu2706}, which solves the Einstein-field Maxwell's equations with an axisymmetric deformation, to determine the stellar properties of magnetic NSs.

The maintenance of the divergenceless constraint ($\nabla. B = 0$) is one of the most difficult problems in the evolution of the relativistic magnetic field equations. Since the LORENE library solves the coupled Einstein-Maxwell field equations allowing for stable magnetized stars, the divergenceless constraint is preserved implying the no-monopole constraint.\\

The maximal-slicing-quasi-isotropic (MSQI) metric is employed for polar spherical symmetry, which allows stars to deform by letting the metric potentials rely on the radial $r$ and angular coordinates  $\theta$ with respect to the magnetic axis. By employing this approach, the field
is produced self-consistently by a macroscopic current, which is a
function of the stellar radius, angle $\theta$ with respect to the symmetry
axis, and dipole magnetic moment $\mu$ for each EoS.  This allows one to control the strength of the magnetic field throughout the star either by the magnetic dipole moment $\mu$ or the dimensionless current
function $f_0$. In this work, we vary the former.

 However, since the TOV equations are still widely used (although incorrectly) to study the stellar properties of stars with any value of the magnetic field, we have also discussed the results from  TOV equations in order to quantify the error the use of TOV introduces in the mass and deformation of stars.

To study the magnetic field effects on the microscopic EoS, we employ a chemical-potential dependent magnetic field, fitted from the solutions of the Einstein-Maxwell's equations. The relation between the magnetic field and the chemical potential depends on the magnetic dipole moment and is given by the relation \citep{DEXHEIMER2017487}
\begin{equation}
B^*(\mu_B)=\frac{(a+b\mu_B +c\mu_B^2)}{B_c^2} \mu\ ,
\end{equation}

where $\mu_B$ is the baryon chemical potential in MeV and $\mu$ is the dipole magnetic moment in units of Am$^2$, so as to produce $B^*$ in units of the electron critical field  $B_c=4.414\times 10^{13}$ G. The coefficients $a$, $b$, and $c$  taken as $a=-0.786$  G$^2$/(Am$^2$), $b=1.24\times 10^{-3}$  G$^2$/(Am$^2$ MeV) and $c=-3.51\times 10^{-7}$  G$^2$/(Am$^2$ MeV$^2$) are obtained from a fit for the magnetic field in the polar direction of a star with a baryon mass of $2.2$ $M_{\sun}$.

\begin{figure}[t!]
	\includegraphics[scale=0.32]{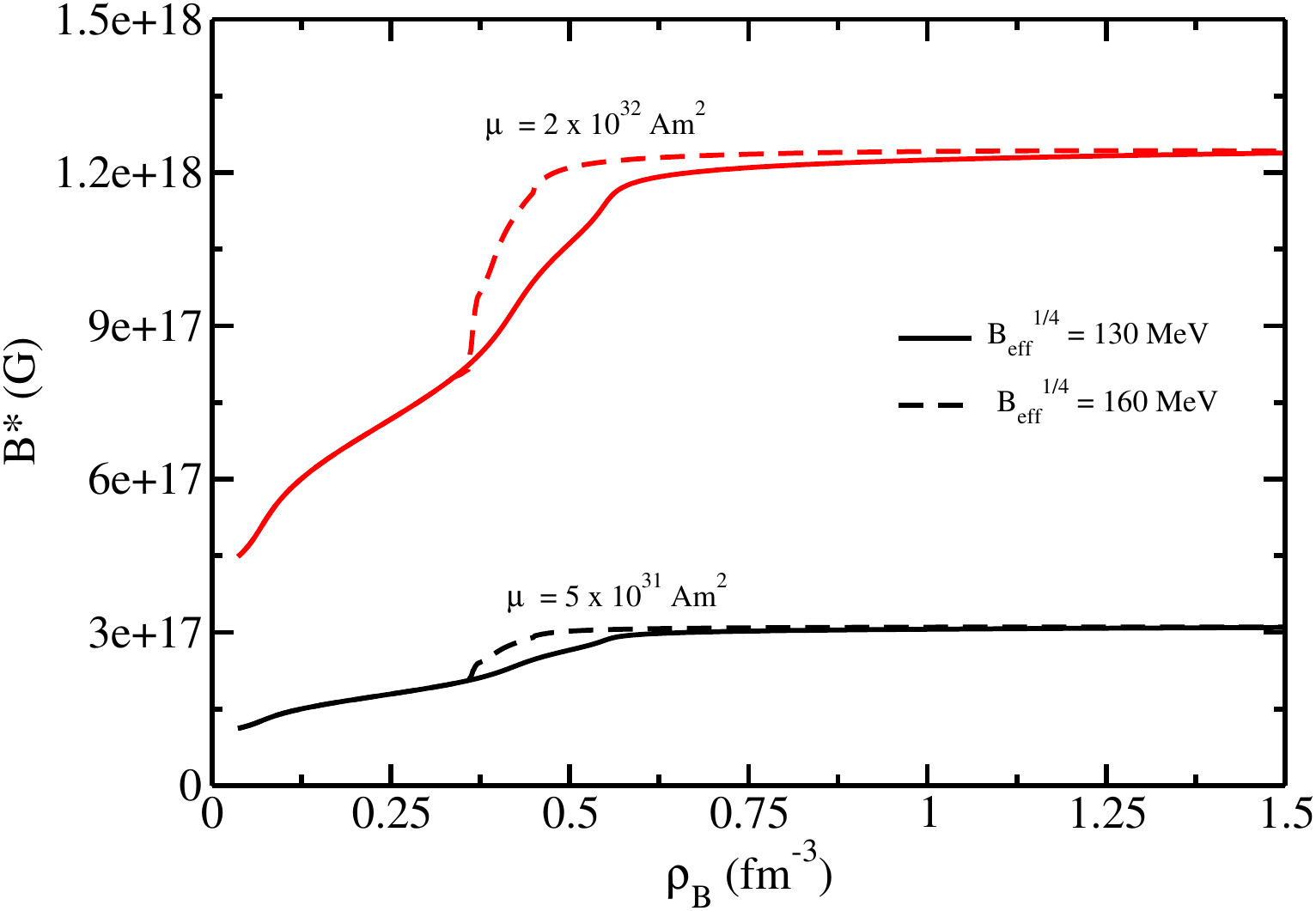}
	\caption{Magnetic field profile as a function of baryon density for the EoSs calculated with different values of effective bag constant. Different values of the magnetic dipole moment are shown.} 
	\label{fig1} 
\end{figure}

Fig. \ref{fig1} shows the magnetic field profile as a function of baryon density obtained for two EoSs with quark deconfinement and mixed phase, one with a larger value of effective bag constant (dashed lines) and one with a smaller value (full lines). They differ because the effective bag constant affects how baryon density and chemical potential relate, with the larger value reproducing a softer quark matter EoS and lower density $\rho_B=\partial P/\partial\mu_B$ (at a given $\mu_B$ or $B^*$) along the mixed phase for a given value of dipole magnetic moment. Regardless, at large densities, a dipole magnetic moment approximately determines the strength of the magnetic field reproduced (see Tab.~\ref{tab1}), reaching either $~10^{17}$ or $~10^{18}$, depending on the dipole magnetic moment. At low densities, the magnetic field is less than one order of magnitude lower than its maximum value, showing a large contrast with ad hoc exponential profiles \citep{PhysRevLett.79.2176,Dexheimer2012-vb,Lopes:2014vva}.

\begin{deluxetable}{p{1.35cm}|p{1.4cm}p{1.4cm}|p{1.4cm}p{1.42cm}}
	\tablenum{1}
	\tablecaption{Magnetic field profile values at low and high densities calculated for EoSs with different effective bag constants and dipole magnetic moments.\label{tab1} }
	\tablewidth{0pt}
	\tablehead{
		&\multicolumn{2}{p{2.7cm}|}{$B^{1/4}_{\rm{eff}} = 130$ MeV} & %
		\multicolumn{2}{p{2.5cm}}{$B^{1/4}_{\rm{eff}} = 160$ MeV}\\
		\hline
		$\mu$ (Am$^2$)& $B_{\rm{low}}$&$B_{\rm{high}}$&$B_{\rm{low}}$&$B_{\rm{high}}$}
	\startdata
	5$\times$10$^{31}$ &$1.1\times10^{17}$&$3\times10^{17}$&$1.1\times10^{17}$&$3.2\times10^{17}$\\
	$2\times10^{32}$
	&$4.5\times10^{17}$&$1.2\times10^{18}$&$4.5\times10^{17}$&$1.3\times10^{18}$\\
	\enddata
\end{deluxetable}

We discuss how high the density can become inside stars depending on dipole magnetic moment and EoS in the next section.\\

\section{Results and Discussion}
\label{sec6}

\begin{figure}
	\includegraphics[scale=0.32]{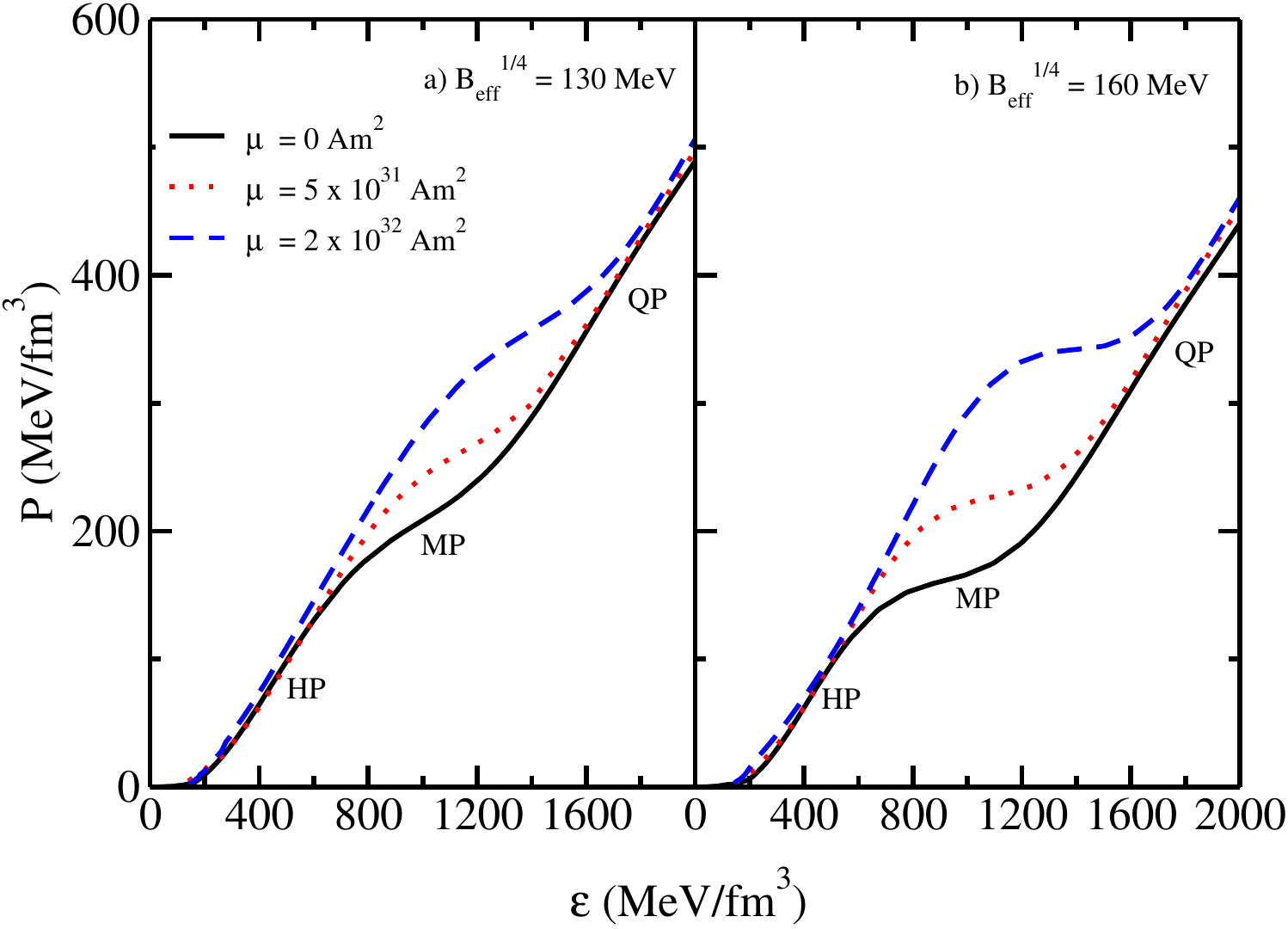}
	\caption{EoS (transverse pressure vs. energy density) for different values of dipole magnetic moment. The left (right) panel shows the EoS with effective bag constant $B^{1/4}_{\rm{eff}}$ = 130 MeV ($B^{1/4}_{\rm{eff}}$ = 160 MeV). The hadronic, mixed, and quark phases are identified, respectively, by HP, MP, and QP. }
	\label{fig2} 
\end{figure}

Fig.~\ref{fig2} displays NS EoSs with quark deconfinement in the presence of the magnetic field profiles previously discussed for different values of the magnetic dipole moment. The left panel shows EoSs with effective bag constant $B^{1/4}_{\rm{eff}} = 130$ MeV, while the right panel shows EoSs with $B^{1/4}_{\rm{eff}} = 160$ MeV. The solid black lines show the EoS at $\mu = 0$ Am$^2$, which corresponds to the zero magnetic field case. All curves include the contribution from matter only (no pure electromagnetic part). 

As already mentioned, a larger value of effective bag constant reproduces a softer quark matter and mixed 
phase EoS (lower pressure). As a consequence, the onset of the mixed and pure quark phases occurs at higher energy densities and densities for the larger value of the effective bag constant \citep{Rabhi_2009,PhysRevC.68.035804}. These statements do not change in the presence of magnetic fields. 

\begin{figure}
\subfloat[]{%
  \includegraphics[clip,width=\columnwidth]{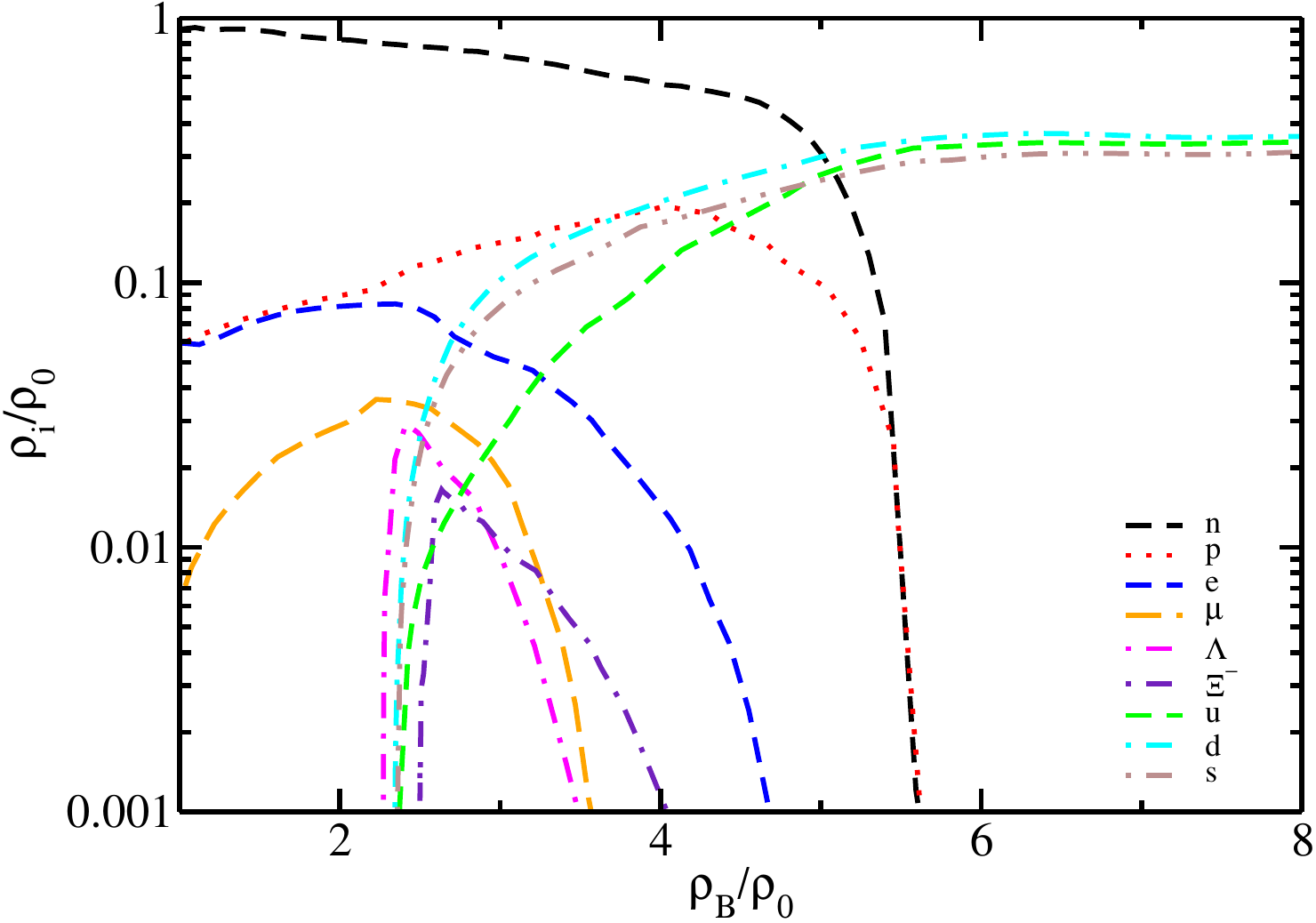}%
}

\subfloat[]{%
  \includegraphics[clip,width=\columnwidth]{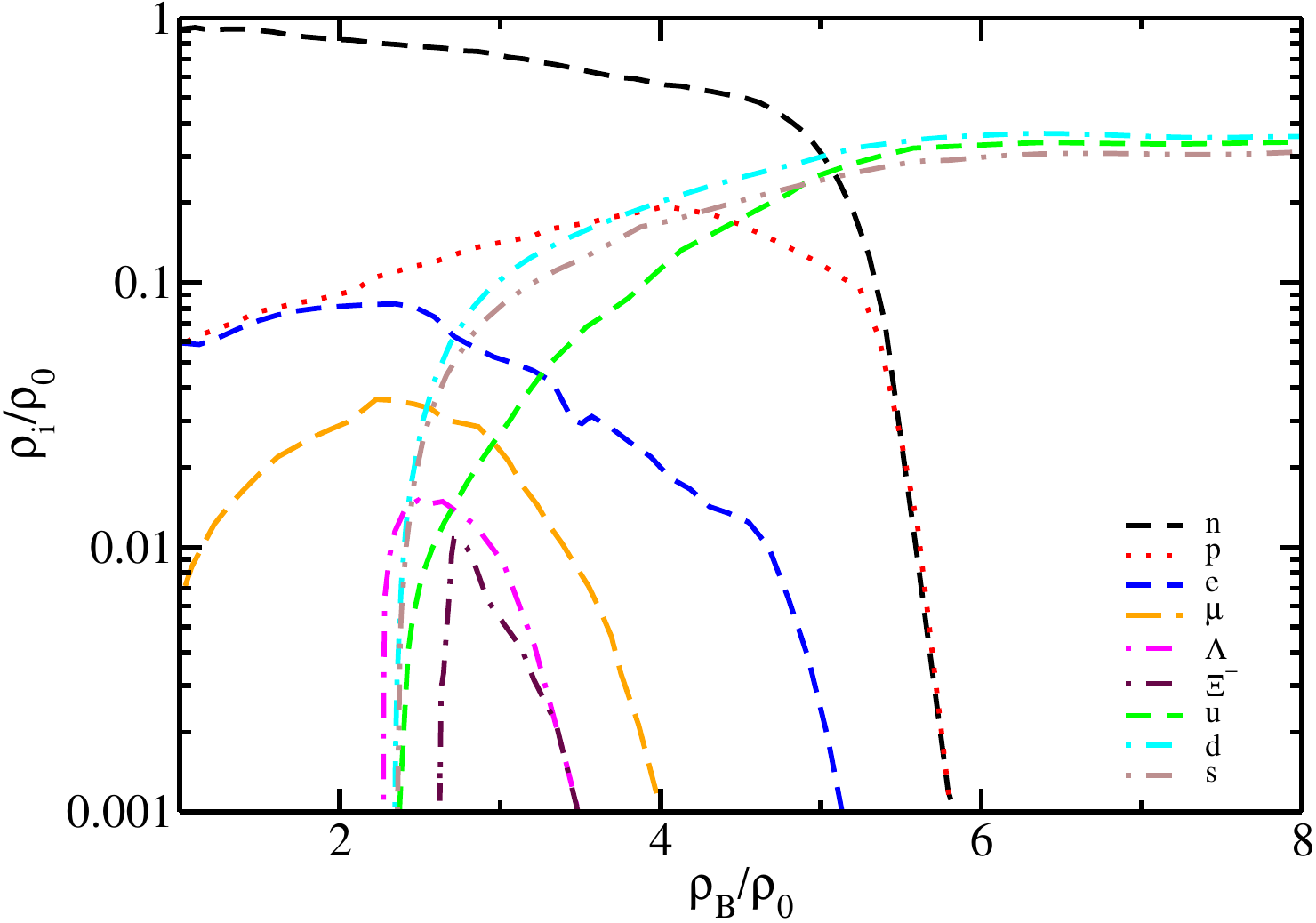}%
}

\subfloat[]{%
  \includegraphics[clip,width=\columnwidth]{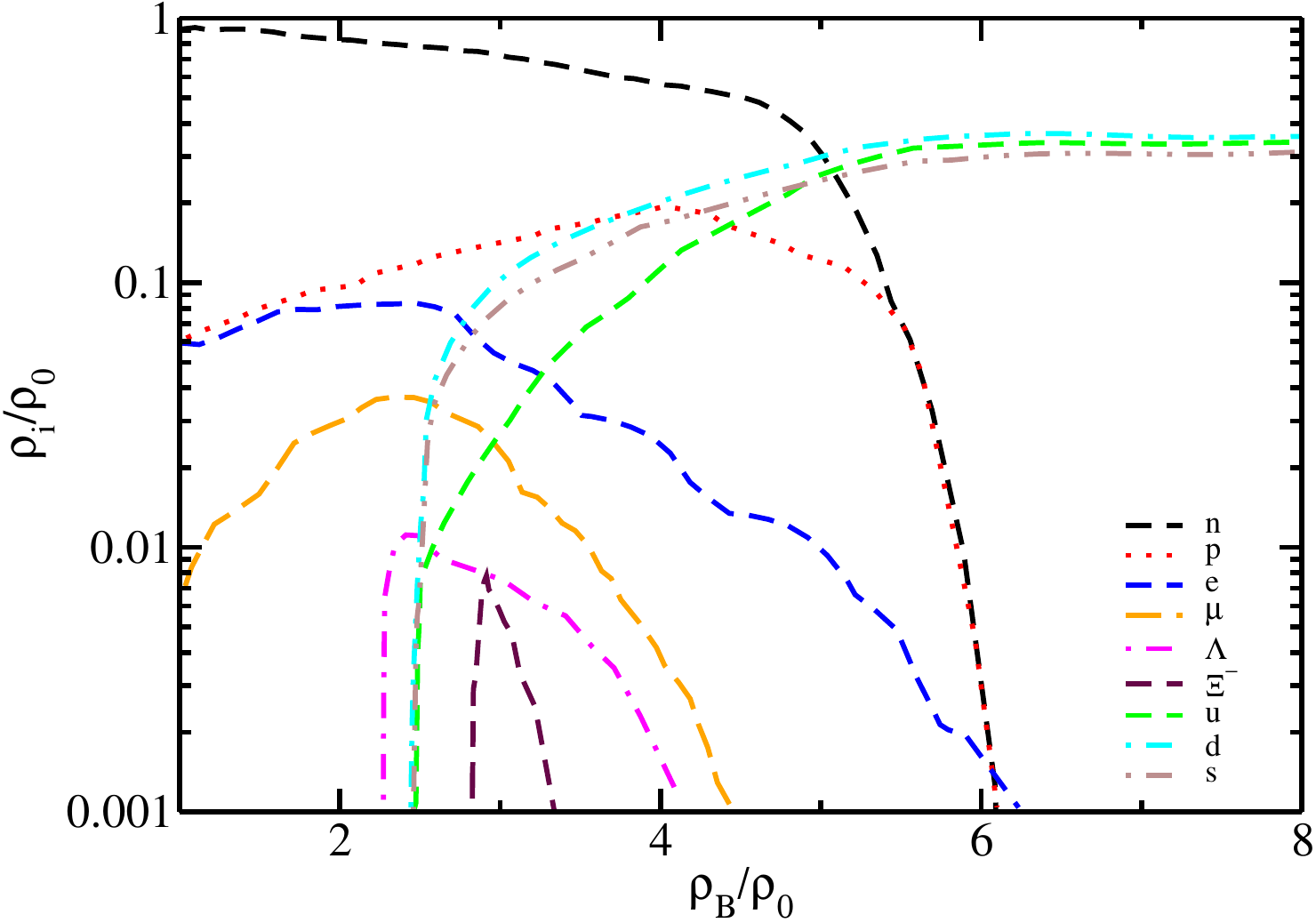}%
}

\caption{Normalized particle fraction of baryons, leptons, and quarks as a function of normalized baryon density for the effective bag constant $B^{1/4}_{\rm{eff}} = 130$ MeV without magnetic field (a) and with magnetic field with different magnetic dipole moments, $\mu$=5$\times$10$^{31}$ Am$^2$ (b) and $\mu$=2$\times$10$^{32}$ Am$^2$ (c).}\label{fig:130}
\end{figure}

For the magnetic dipole moment of $\mu = 5\times 10^{31}$ Am$^2$, which corresponds to a high-density magnetic field of $\sim10^{17}$ G, for the lower effective bag constant  the hadronic phase in the low-density region is not affected by the magnetic field, while the pure quark phase in the high-density is slightly stiffer than the EoS without magnetic field. The mixed phase region extends to higher densities. Similar behavior is observed for the EoSs with a higher effective bag constant. 
For the magnetic dipole moment of $\mu = 2\times 10^{32}$ Am$^2$, which corresponds to a high-density magnetic field of $\sim10^{18}$ G, both the hadronic phase and pure quark phases become stiffer. The mixed phase region is even broader (than with lower $\mu$).
The pure quark phase appears at very large densities, and hence, will occupy a small part (if any) of NSs. For the larger effective bag constant, the onset of mixed phase and pure quark phase is strongly affected. This is because at the densities at which it occurs the magnetic field is much stronger. Thus, we see that the transitions to the mixed phase and to the pure quark phase are affected by the inclusion of the magnetic field and depend upon its strength. See Tab.~\ref{tab2} for the exact density of the phase boundaries.

\begin{figure}[t!]
\subfloat[]{%
  \includegraphics[clip,width=\columnwidth]{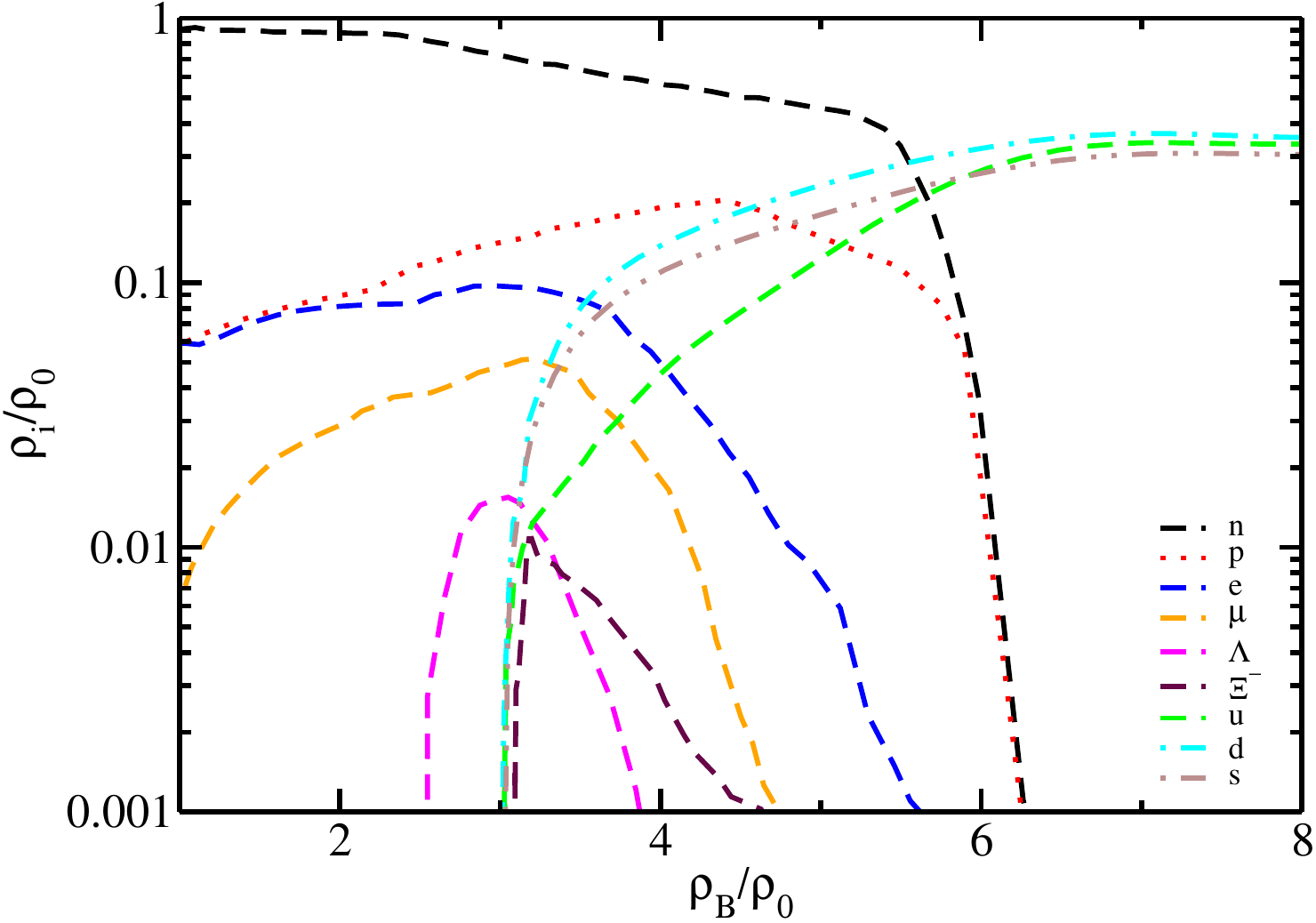}%
}

\subfloat[]{%
  \includegraphics[clip,width=\columnwidth]{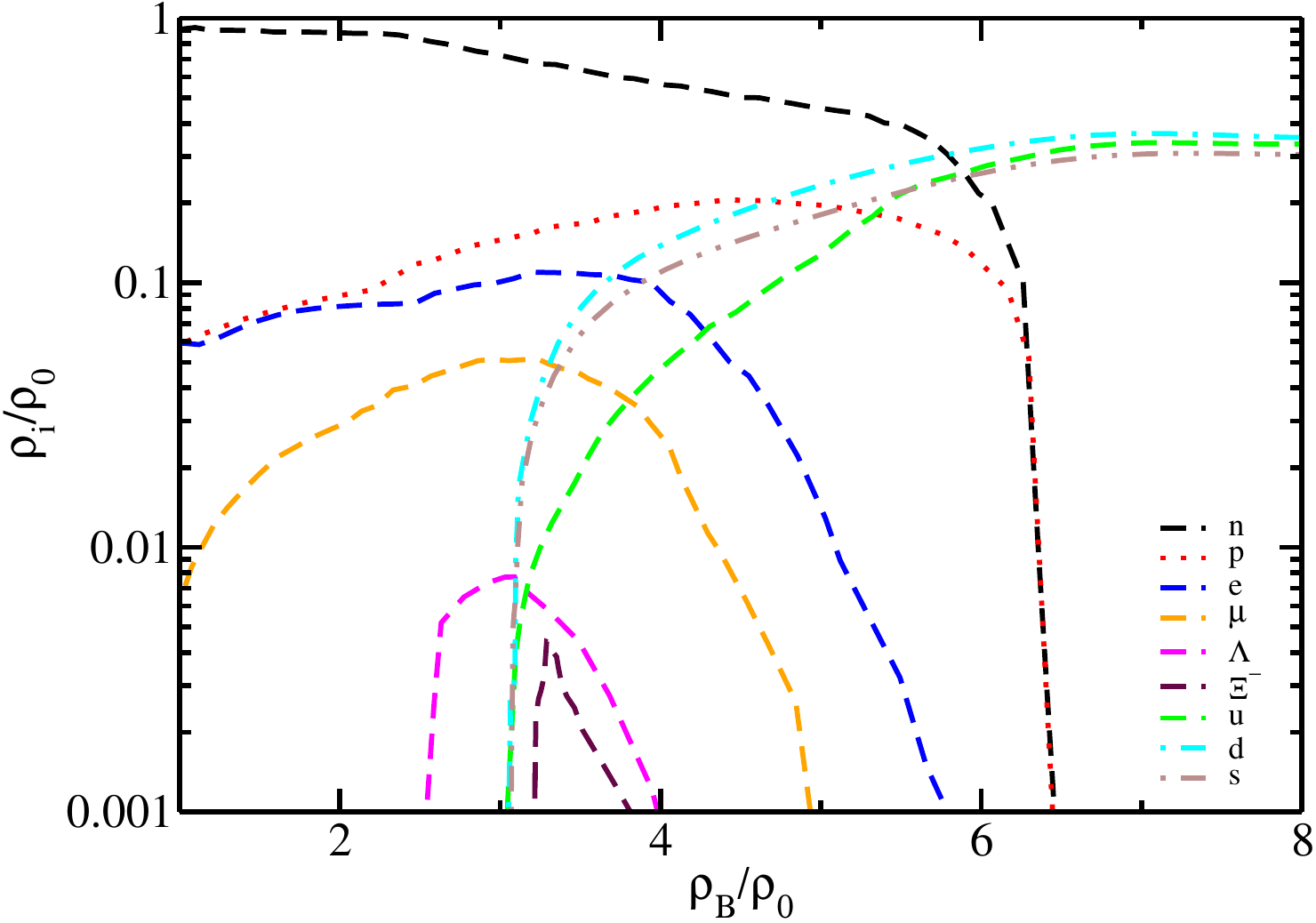}%
}

\subfloat[]{%
  \includegraphics[clip,width=\columnwidth]{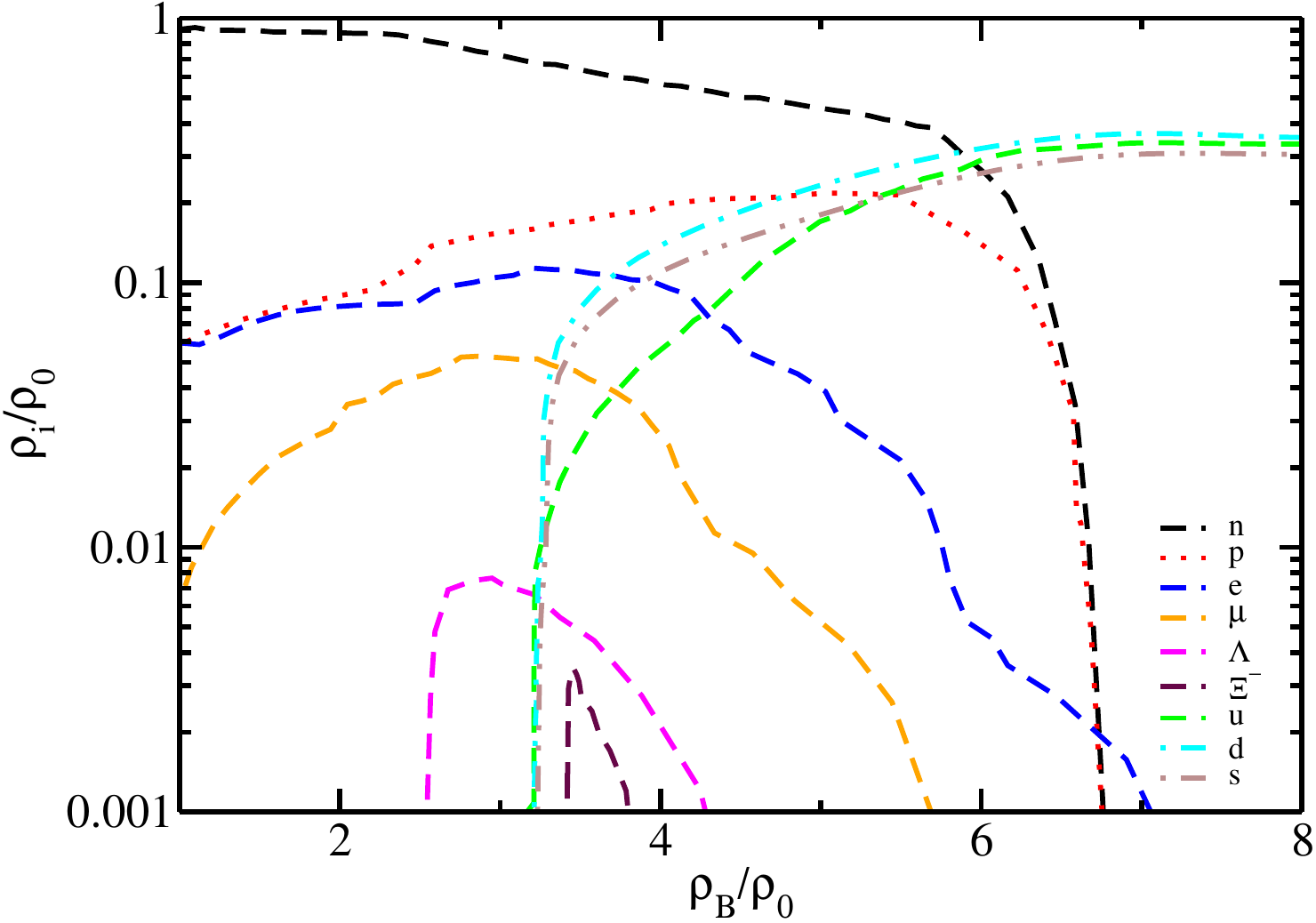}%
}
\caption{Same as Fig. \ref{fig:130} but with effective bag constant $B^{1/4}_{\rm{eff}}$ = 160 MeV.}\label{fig:160}
\end{figure}

\begin{figure}[]
	\includegraphics[scale=0.35]{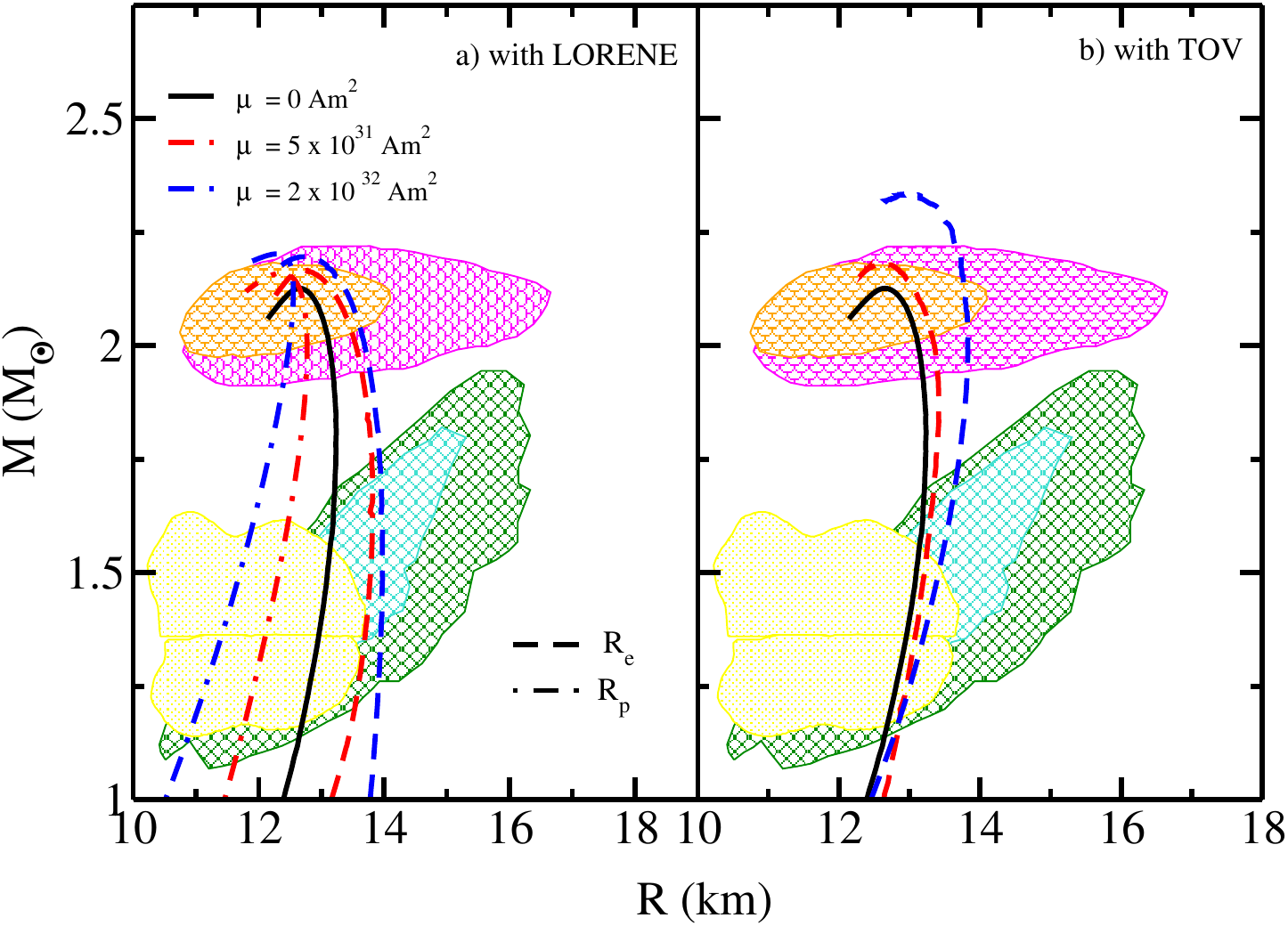}
	\caption{Mass-radius diagram for hybrid stars with effective bag constant $B^{1/4}_{\rm{eff}}$ = 130 shown for different values of magnetic dipole moment. The left panel shows results obtained using the LORENE library, while the right panel shows solutions from TOV.  $R_e$ and $R_p$ represent the equatorial and polar radii of the star. Recent constraints on mass and radius are also shown \citep{Abbott_2020a,Demorest2010,Antoniadis1233232,Cromartie2020,Miller_2019a,Riley_2019,Miller_2021,Riley:2021pdl}.} 
	\label{figmr130} 
\end{figure}

Fig.~\ref{fig:130} shows the normalized particle population of baryons, leptons, and quarks without and with magnetic field effects with different values of the magnetic dipole moment. These plots correspond to the EoSs obtained with the effective bag constant $B^{1/4}_{\rm{eff}} = 130$ MeV. For $\mu = 0$ Am$^2$, the hadrons disappear in the mixed phase and quarks appear smoothly. Even though hyperons are included in the calculations, they are partially suppressed by the appearance of the quark phase, a small amount of $\Lambda$ hyperons appears at density around $2.2$ $\rho_{0}$, just before the hadron-quark mixed phase starts, and a small amount of $\Sigma^-$ appears in the mixed phase. The density of leptons, $e^-$ and $\mu^-$, is significant in the hadron phase but vanishes in the mixed phase. Since the quarks ($d$ and $s$) are negatively charged, there is no necessity for leptons in the quark phase to maintain beta equilibrium and charge neutrality conditions.
For the lower value of the magnetic dipole moment (when compared with $\mu=0$), oscillations appear in the particle population of charged particles, which arise due to the Landau levels. With the increasing density, charged particles (especially leptons) are enhanced, as seen in the Fig.~\ref{fig:130}(b). As already discussed, the density at which the mixed phase appears and the density range of the mixed phase increase with larger values of $\mu$. The hyperons are suppressed due to an increase in the proton density \citep{Broderick:2001qw}. These effects are enhanced in magnitude for the larger value of the magnetic dipole moment, as seen in Fig.~\ref{fig:130}(c).

\begin{figure}
	\includegraphics[scale=0.35]{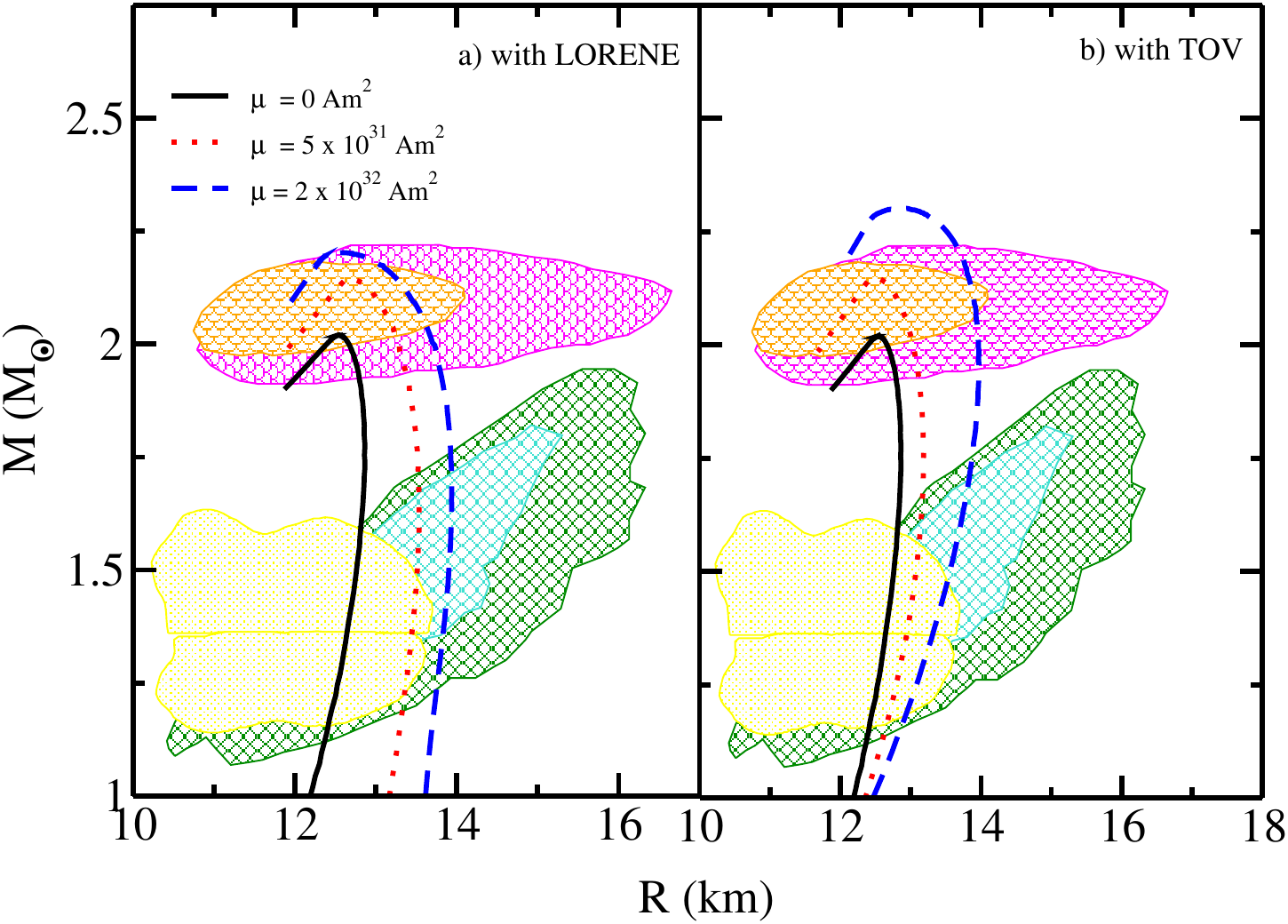}
	\caption{ Same as Fig. \ref{figmr130} but showing only the equatorial radius for $B^{1/4}_{\rm{eff}}$ = 160 MeV.}
	\label{figmr160} 
\end{figure}

Fig.~\ref{fig:160} also shows the populations of baryons, leptons, and quarks but now with a larger effective bag constant $B^{1/4}_{\rm{eff}}$ = 160 MeV. Similar to Fig.~\ref{fig:130}, some of the hyperons appear just before the onset of mixed phase and some after, all being suppressed for larger dipole magnetic moments. As the magnetic dipole moment increases, the lepton fraction is enhanced to the point that a small number of leptons extends into the pure quark phase. For higher values of the magnetic dipole moment, the mixed phase region extends up to 7$\rho_{0}$. In the mixed phase, the abundance of $u$-quarks is found to be enhanced, while those of $d$ and $s$ quarks remain practically the same.\\
\begin{deluxetable}{p{1.35cm}|p{1.4cm}p{1.4cm}|p{1.4cm}p{1.42cm}}[t!]
	\tablenum{2}
	\tablecaption{Normalized baryon density phase boundaries: The Beginning and End of the mixed phase for the EoSs calculated with different effective bag constants and dipole magnetic moments. \label{tab2}}
	\tablewidth{0pt}
	\tablehead{
		&\multicolumn{2}{p{2.7cm}|}{$B^{1/4}_{\rm{eff}} = 130$ MeV} & %
		\multicolumn{2}{p{2.5cm}}{$B^{1/4}_{\rm{eff}} = 160$ MeV}\\
		\hline
		$\mu$ (Am$^2$)& $\rho_{\rm{beg}}/\rho_0$ &$\rho_{\rm{end}}/\rho_0$&$\rho_{\rm{beg}}/\rho_0$&$\rho_{\rm{end}}/\rho_0$ }
	\startdata
	$0$&$2.32$&$5.63$&$3.04$&$6.27$\\
	$5\times10^{31}$&$2.35$&$5.78$&$3.08$&$6.38$ \\ 
	$2\times10^{32}$&$2.47$&$6.07$&$3.21$&$6.75$\\
	\enddata
\end{deluxetable}
Fig.~\ref{figmr130} shows the mass-radius diagram for the hybrid stars obtained for the effective bag constant $B^{1/4}_{\rm{eff}} = 130$ MeV without a magnetic field and with magnetic field calculated for different values of the magnetic dipole moment. The left panel shows the results obtained from solving Einstein and Maxwell's equations with an axisymmetric deformation (LORENE library), while the right panel displays the results using solutions from spherically symmetric TOV equations. For $\mu=B=0$ they coincide. Without a magnetic field, the maximum mass produced is $2.13$ $M_{\sun}$ at a radius of 12.6 km. The radius of the canonical mass, $1.4$ $M_{\sun}$ is around 13 km, which satisfies the radius constraints from PSR J0030+0451 by NICER \citep{Miller_2019a,Riley_2019}. The recent NICER constraint on the radius of the $2$ $M_{\sun}$ pulsar J0740+6620 NS is also satisfied \citep{Miller_2021,Riley:2021pdl}. Maximum mass constraints and constraints from LIGO/VIRGO are also satisfied \citep{PhysRevLett.119.161101,PhysRevLett.121.161101}.\\

 For the results obtained using the LORENE library, the deformation present due to the poloidal magnetic field makes the star oblate. Because of the deformation present, the equatorial radius, $R_e$, and the polar radius,  $R_p$, of the star are different and change with the increase in the magnetic field strength. Since the TOV equations are used for the spherically symmetric stars, the equatorial and the polar radii are equal and remain so in the vicinity of the magnetic field.

\begin{figure}[t!]
	\includegraphics[scale=0.35]{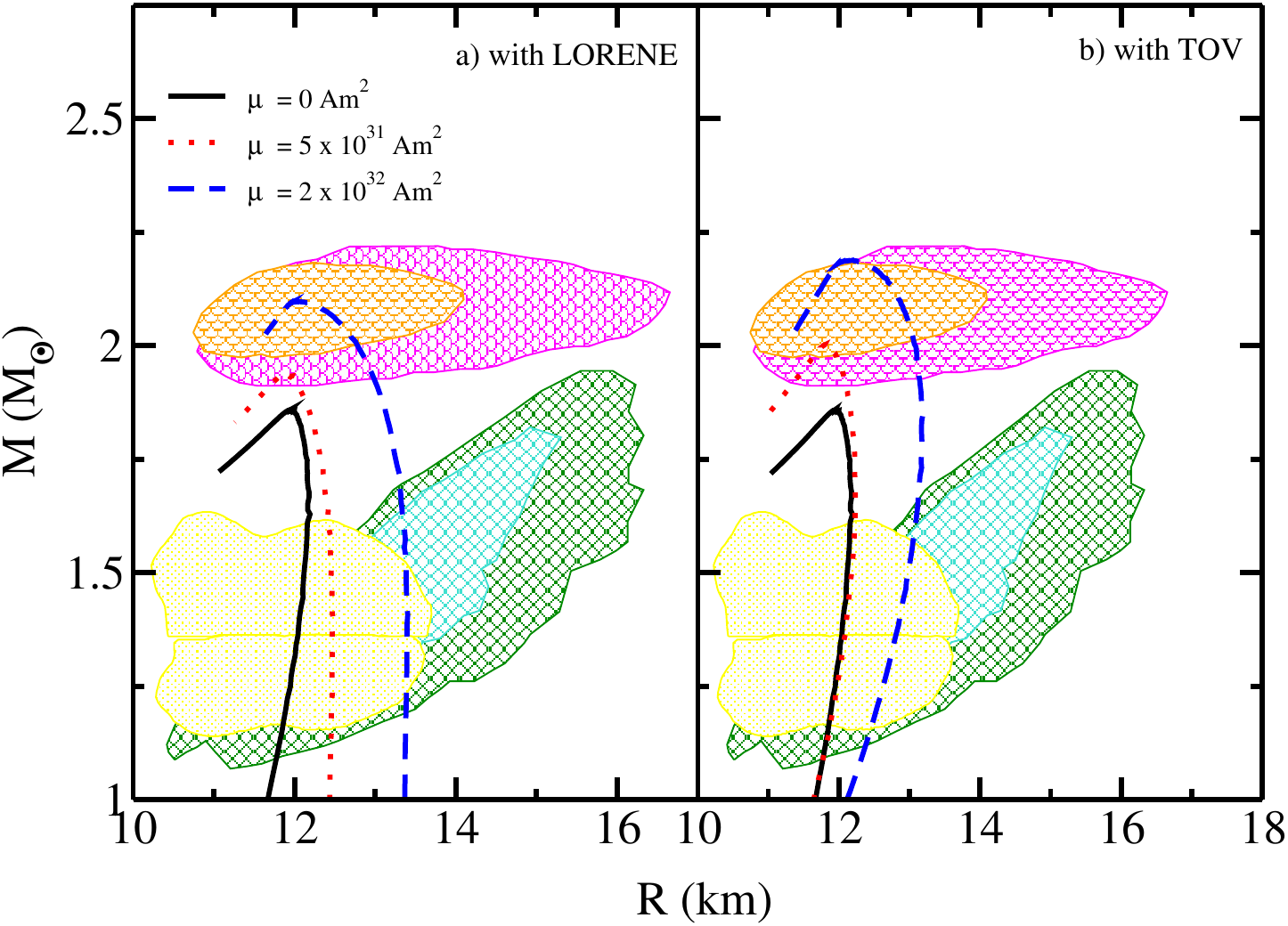}
	\caption{ Same as Fig. \ref{figmr130} but showing only the equatorial radius for $B^{1/4}_{\rm{eff}}$ = 180 MeV.} 
	\label{figmr180} 
\end{figure}

With increasing magnetic dipole moment, the mass and radius of entire stellar sequences increase. For $\mu = 5\times 10^{31}$ Am$^2$, which corresponds to a  magnetic field of 3 $\times$ 10$^{17}$ G in the center of the maximum-mass star, the equatorial radius of low-/intermediate-mass stars obtained using the LORENE library differs significantly from the radius obtained from TOV, $0.6$ km for the canonical mass when compared with $0.1$ km for the maximum mass. For $\mu$ = 2 $\times$ 10$^{32}$ Am$^2$, which corresponds to a  magnetic field of 1.2 $\times$ 10$^{18}$ G in the center of the maximum-mass star, both the radius and mass of all stars from the sequence obtained using LORENE differ significantly from the values obtained from TOV. The difference (when compared to the equatorial radius) is $0.7$ km for the canonical mass and $-0.2$ km for the maximum mass (with the minus meaning that now TOV gives the larger radius). In the case of TOV, because of the required spherical symmetry, the absence of the Lorentz force does not enlarge the equator of less massive stars that possess a softer EoS, instead, the excess magnetic energy that would deform the star is added to the mass, which becomes unphysically large. 
For the results with the LORENE library, we can see that at the maximum mass of the star, the equatorial and the polar radius almost overlap each other, indicating almost no deformation of the star.

Fig.~\ref{figmr160} also displays the mass-equatorial radius profiles for hybrid stars but with effective bag constant $B^{1/4}_{\rm{eff}}$ = 160 MeV. For $\mu$ = 0 Am$^2$, the maximum mass obtained lowers to $2.02$ $M_{\sun}$ with corresponding radius of $12.5$ km (when compared with Fig.~\ref{figmr130}). The equatorial radius at the canonical mass is $12.68$ km. Our results are still in agreement with NICER, LIGO/VIRGO, and mass constraints. Using LORENE, as the magnetic dipole moment increases, the maximum mass increases more than for the previous effective bag constant considered. The equatorial radius of the whole sequence also increases more. 
The solutions from TOV, do not depend as much on the effective bag constant, and in this case, are farther from reproducing low-/intermediate mass results from LORENE. 
A difference of around $0.4$ km is observed in the radius at the canonical mass and $-0.3$ for the maximum mass with measurements from LORENE and TOV. 
This implies that the difference between the two approaches depends on the EoS (particle composition and interactions) and the mass of the star we are calculating the deformation for.

Table \ref{tab4} shows the maximum mass, corresponding radius, canonical mass radius, and canonical mass dimensionless tidal deformability obtained at different values of the magnetic dipole moment using both the LORENE library and TOV solutions. For the LORENE library, the radius shown in the table corresponds to the equatorial radius.
Without a magnetic field, the dimensionless tidal deformability for the canonical mass satisfies the constraints from  GW170817, $\tilde{\Lambda}<800$ ($B^{1/4}_{\rm{eff}} = 160$ MeV) and the recently observed GW190814 data, $\Lambda_{1.4} = 616_{-158}^{+273}$ ($B^{1/4}_{\rm{eff}} = 130$ and $160$ MeV). This constraint from GW190814 data is obtained by considering its secondary component to a supermassive NS \citep{Abbott_2020a}. 

To further verify our results, we calculate the deconfinement phase transition using the effective bag constant of $B^{1/4}_{\rm{eff}} = 180$ MeV. Both the publicly available LORENE and spherically symmetric TOV equations are used to produce the mass-radius profiles and analyze the differences in the stellar properties. Fig.~\ref{figmr180} displays the mass-equatorial radius diagram for 180 MeV effective bag constant for different values of the magnetic dipole moment. For $\mu = 0$ Am$^2$ case, the maximum mass and the canonical mass radius for the hybrid star configuration obtained is $1.86$ M$_{\sun}$ and $11.94$ km, respectively. The canonical radius is $12.07$ km. In this case, the maximum mass is not high enough to fulfill modern mass constraints.
For larger dipole magnetic moments, there is a large increase in mass (than the other effective bag constants) using LORENE and the radii, particularly for low-/intermediate mass stars, become very large. This reinforces the results we discussed above. 
\begin{deluxetable}{p{1.35cm}|p{1.4cm}p{1.4cm}|p{1.4cm}p{1.42cm}}
	\tablenum{3}
	\tablecaption{Maximum Mass, Maximum mass Equatorial Radius, Canonical mass radius, and Canonical mass'	dimensionless tidal deformability of hybrid stars obtained with different effective bag constants and magnetic dipole moments.\label{tab4}}
	\tablewidth{0pt}
	\tablehead{
			&\multicolumn{4}{c}{Using LORENE and $B^{1/4}_{\rm{eff}} = 130$ MeV}\\  %
		\hline
		$\mu$ (Am$^2$)&$M_{max}$&$R_{max}$&$R_{1.4}$&$\Lambda_{1.4}$}
		\startdata
		0&2.13&12.64&13.01&879.54\\
		5$\times$10$^{31}$&2.16&12.67&13.73&1223.98\\ 
		2$\times$10$^{32}$&2.20&12.78&13.96&1276.95\\
		\hline
		& \multicolumn{4}{c}{Using TOV and $B^{1/4}_{\rm{eff}} = 130$ MeV}\\
		\hline
		0&2.13&12.64&13.01&879.54\\
		5$\times$10$^{31}$&2.18&12.66&13.13&956.02\\ 
		2$\times$10$^{32}$&2.33&12.98&13.26&1236.42\\
		\hline
		&\multicolumn{4}{c}{Using LORENE and $B^{1/4}_{\rm{eff}} = 160$ MeV}\\  %
		\hline
		0&2.02&12.53&12.68&700.94\\
		5$\times$10$^{31}$&2.10&12.64&13.46&946.38\\ 
		2$\times$10$^{32}$&2.19&12.56&13.86&1268.99\\
		\hline
		& \multicolumn{4}{c}{Using TOV and $B^{1/4}_{\rm{eff}} = 160$ MeV}\\
		\hline
		0&2.02&12.53&12.68&700.94\\
		5$\times$10$^{31}$&2.15&12.56&12.97&923.13\\ 
		2$\times$10$^{32}$&2.29&12.89&13.39&1194.26\\
		\hline
		&\multicolumn{4}{c}{Using LORENE and $B^{1/4}_{\rm{eff}} = 180$ MeV}\\  %
		\hline
		0&1.86&11.94&12.07&472.26\\
		5$\times$10$^{31}$&1.94&11.87&12.47&567.22\\ 
		2$\times$10$^{32}$&2.08&12.07&13.27&976.83\\
		\hline
		& \multicolumn{4}{c}{Using TOV and $B^{1/4}_{\rm{eff}} = 180$ MeV}\\
		\hline
		0&1.86&11.94&12.07&472.26\\
		5$\times$10$^{31}$&2.01&11.82&12.10&532.06\\ 
		2$\times$10$^{32}$&2.19&12.19&12.92&872.23\\
	\enddata
\end{deluxetable}

\section{Summary and Conclusions}
\label{summary}
We study the effect of strong magnetic fields on the phase transition between baryons and quarks in the core of NSs. For this purpose, we use the widely known DD-RMF model with the DD-MEX parameter set, which offers enough flexibility to meet both nuclear and astrophysical constraints. The hyperon couplings are determined from SU(6) symmetry. For the quark phase, we use the vBag, with the construction of a mixed phase, which allows describing strange hybrid NSs that achieve the $2$ M$_{\sun}$ limit. Several values of the effective bag constant, $B^{1/4}_{\rm{eff}}=130, 160, 180$ MeV, are considered. 

To study magnetic field effects on the EoS, a physically motivated profile is applied, which depends quadratically on the chemical potential. As a consequence, the profile as a function of density depends on the EoS around the phase transition, and it is different for different effective bag constants.
At large densities, the magnetic field approximately only depends on the magnetic dipole moment, reaching $10^{18}$ G for the largest magnetic dipole moment studied, $\mu = 2\times 10^{32}$ Am$^2$. Overall, the EoS stiffens in the presence of magnetic fields, but the effects are stronger in the mixed phase, which is also wider (in density and energy density) and takes place at larger densities. The effects are stronger for larger magnetic dipole moments and effective bag constants. The hyperons, which appear around the onset of the mixed phase are suppressed by the magnetic field (due to an increased proton density). Oscillations appear in the particle population of charged particles, which arise due to the Landau levels. Overall, the population of charged particles (especially leptons) is enhanced. The population of the $u$-quark is enhanced, when compared with the other quarks.

The LORENE library is used to calculate stellar properties of NSs that present strong magnetic fields, and are therefore not spherically symmetric. However in order to quantify the anisotropy, we also use solutions from the spherically symmetric TOV equations, and discuss the difference. For $\mu=B=0$ they of course coincide. In this case, the maximum mass produced with effective bag constants $B^{1/4}_{\rm{eff}}=130$ and $160$ MeV, $2.13$ and $2.02$ M$_{\sun}$, respectively, satisfy the mass and radius constraints from various measurements. 

Using the LORENE library, the poloidal magnetic field turns NSs oblate, increasing the equatorial radius. While the mass and radius of all stars of a family are modified by the magnetic field, this is not the case for TOV, in which case just massive stars become larger and their masses become unphysically large. This qualitative behavior is independent of the effective bag constant. Quantitatively, the effects of  the magnetic field have a greater effect on the mass and radii of the entire family of stars described by lower effective bag constant (softer EoS) using LORENE. Using TOV, the effects of the magnetic field do not depend as much on the effective bag constant. This means that, as discussed in detail for the first time in this paper, the difference between anisotropic and isotropic general-relativity solutions depends on the EoS and mass of the star we are calculating the deformation for. 

Interestingly, for some fixed stellar masses, the difference in radius between LORENE and TOV is larger for lower values of $\mu$ and $B$. This points to the fact that TOV should not be used to study strongly magnetized stars even for lower values of dipole magnetic moment. Overall, the canonical radius is always smaller for TOV, and the difference increases with effective bag constant. For the radius of the maximum-mass star, TOV can have smaller (lower $\mu$) or larger (larger $\mu$) radius than LORENE. The maximum mass of the sequence is always too large for TOV, the difference increasing with $\mu$ and effective bag constant. In addition, as TOV does not involve the solution of Maxwell's equations in the presence of magnetic fields, it implies the erroneous existence of a magnetic monopole.

Future extensions of this work include adding a crust (see Ref.~\citep{Wang:2022sxx} for a discussion of magnetic field effects on the neutron-star crust), the effects of temperature, fast rotation, and anomalous magnetic moment (AMM) of baryons, as well as exploring different parametrizations of the models that allow for smoother phase transitions. We also intend to use our EoSs in numerical relativity simulations of merging binary NSs, as these are the key to learning about the physical processes involved in events such as GW170817. Magnetic fields, in particular, are known to play an important role in post-merger evolution, fueling relativistic jets, and shaping electromagnetic counterpart signals \citep{Ciolfi:2013dta,Giacomazzo:2014qba,Ciolfi2020,Palenzuela:2021gdo,Ruiz:2021qmm}. 
\section{Acknowledgment}
I.A.R. and V.D. are thankful to T. Kl\"{a}hn and H. Pais for the discussion on the quark matter and magnetic EoS. A.A.U. acknowledges the Inter-University Centre for Astronomy and Astrophysics, Pune, India for support via an associateship and for hospitality. V. D. acknowledges support from the National Science Foundation under grants PHY1748621, MUSES OAC-2103680, and NP3M PHY-2116686. I.L. and I.A.R. also acknowledge the Funda\c c\~ao para a Ci\^encia e Tecnologia (FCT), Portugal,
for the financial support to the Center for Astrophysics and Gravitation (CENTRA/IST/ULisboa)
through the grant Project~No.~UIDB/00099/2020  and grant No. PTDC/FIS-AST/28920/2017.
\appendix
\section{}
\label{A}
\numberwithin{equation}{section}
Within the relativistic density-dependent mean field (DD-RMF) model, the equation of motion for baryons obtained by applying the Euler Lagrange equations to the Lagrangian density (Eq.~\ref{lag}) is written as
\begin{align}
\sum_{b}\Bigg[i\gamma_{\mu}D^{\mu}-\gamma^0 \Bigg(g_{\omega}(\rho_b)\omega &+\frac{1}{2}g_{\rho}(\rho_b)\rho \tau_3 +\sum_R (\rho_b)\Bigg)-M_{b}^*\Bigg]\psi_{b} =0\ ,
\end{align} 
where the rearrangement term due to the density dependence of the coupling constants is
\begin{equation}
\sum_R(\rho_b) = -\frac{\partial g_{\sigma}}{\partial \rho_b}\sigma \rho_s +\frac{\partial g_{\omega}}{\partial \rho_b}\omega \rho_b+\frac{1}{2}\frac{\partial g_{\rho}}{\partial \rho_b}\rho \rho_3\ .
\end{equation}
The equations of motion for the meson fields are
\begin{align}
m_{\sigma}^2 \sigma &= g_{\sigma}(\rho_b)\rho_s\ , \nonumber \\
m_{\omega}^2 \omega &= g_{\omega}(\rho_b)\rho_b\ , \nonumber  \\
m_{\rho}^2 \rho &= \frac{g_{\rho}(\rho_b)}{2}\rho_3\ ,
\end{align} 
where the total scalar density $\rho_s$, baryon (vector) density $\rho_b$, and isovector density $\rho_3$ are given as
\begin{align}
\rho_s &= \sum_{b}\bar{\psi}\psi=\sum_{b}\frac{1}{2\pi^2}\left(k_b M^*E^*-{M^*}^3\ln{\frac{k_b+E^*}{M^*}}\right)\ , \nonumber \\
\rho_b &= \sum_{b}\psi^{\dagger}\psi =\sum_{b} \frac{k_b^3}{3\pi^2} \ , \nonumber\\ 
\rho_3 &= \sum_{b}\psi^{\dagger}\tau_3\psi =\rho_p -\rho_n +2\rho_{\Sigma^+}-2\rho_{\Sigma^-}-3\rho_{\Xi^0}-\rho_{\Xi^0}\ \ , \nonumber\\ 
\rho_l &=\sum_{l}\psi^{\dagger}\psi =\sum_{l} \frac{k_l^3}{3\pi^2} \ .
\end{align}
With the above equations, the expressions for the energy density and pressure are
%
%
%
\begin{align} \label{eq18}
\mathcal{E}_m &= \sum_{b} \frac{1}{\pi^2}\Bigg[ \Bigg(\frac{1}{4}k_b^3 +\frac{1}{8} (M_b^*)^2 k_b \Bigg) E_b^* - 
\frac{1}{8} (M_b^*)^4 \ln \frac{k_{b}+E_b^*} {M_b^*} \Bigg] + \sum_{l} \frac{1}{\pi^2}\Bigg[ \Bigg(\frac{1}{4}k_l^3 +\frac{1}{8} (m_l)^2 k_l \Bigg) E_l  \nonumber \\
&-\frac{1}{8} (m_l^*)^4 \ln \frac{k_{l}+E_l} {m_l} \Bigg]
+\frac{1}{2}m_{\sigma}^2 \sigma^2+\frac{1}{2}m_{\omega}^2 \omega^2 +\frac{1}{2}m_{\rho}^2 \rho^2+g_{\omega}(\rho_b)\omega \rho_b+\frac{g_{\rho}(\rho_B)}{2}\rho \rho_3\ ,  \nonumber \\
P_m&= \sum_{b} \frac{1}{3\pi^2}\Bigg[ \Bigg(\frac{1}{4}k_b^3 -\frac{3}{8} (M_b^*)^2 k_b \Bigg) E_b^* + 
\frac{3}{8} (M_b^*)^4 \ln \frac{k_{b}+E_b^*} {M_b^*} \Bigg] + \sum_{l} \frac{1}{3\pi^2}\Bigg[ \Bigg(\frac{1}{4}k_l^3 -\frac{3}{8} (m_l)^2 k_l \Bigg) E_l  \nonumber \\
&+\frac{3}{8} (m_l)^4 \ln \frac{k_{l}+E_l} {m_l} \Bigg]
-\frac{1}{2}m_{\sigma}^2 \sigma^2+\frac{1}{2}m_{\omega}^2 \omega^2 +\frac{1}{2}m_{\rho}^2 \rho^2-\rho_b \sum_R (\rho_b)\  ,
\end{align}
where $E_{b}^*=\sqrt{k_{b}^2+M_b^{*2}}$ and $E_{l}=\sqrt{k_{l}^2+m_l^{2}}$.
The rearrangement term $\sum_{R}(\rho_b)$ contributes to the pressure only.\\
The expressions for the total charged baryon, uncharged baryon, and lepton energy densities in the presence of a magnetic field become
\begin{align}
\mathcal{E}_{cb}&= \frac{|q_{cb}|B}{4\pi^2}\sum_{\nu=0}^{\nu_{max}}r_{\nu} %
\times \Bigg[k_{F}^{cb}E_F^{cb}+(M_{cb}^{*2}+2\nu|q_{b}|B) \ln\Bigg(\frac{k_{F}^{cb}+E_F^{cb}}{\sqrt{M_{cb}^{*2}+2\nu|q_{cb}|B}}\Bigg)\Bigg]\ , \nonumber \\
\mathcal{E}_{ub}&= \frac{1}{8\pi^2}\Bigg[k_{F}^{ub}(E_F^{ub})^3+(k_F^{ub})^3E_F^{ub} -M_{ub}^{*4} \ln \Bigg(\frac{k_{F}^{ub}+E_F^{ub}}{M_{ub}^*}\Bigg)\Bigg]\  ,\nonumber \\
\mathcal{E}_{l}&= \frac{|q_l|B}{4\pi^2}\sum_{\nu=0}^{\nu_{max}}r_{\nu} \times \Bigg[k_{F}^{l}E_F^{l}+(m_{l}^{2}+2\nu|q_l|B) \ln\Bigg(\frac{k_{F}^{l}+E_F^{l}}{\sqrt{m_{l}^2+2\nu|q_l|B}}\Bigg)\Bigg]\ .
\end{align}
The scalar and vector density for a given charged baryon $cb$, uncharged baryon $ub$, and lepton is as follow \citep{Broderick_2000}\\
\begin{align}
\rho_s^{cb}&=\frac{|q_{cb}|B M_{cb}^{*2}}{2\pi^2}\sum_{\nu=0}^{\nu_{max}}r_{\nu}  \ln\Bigg(\frac{k_{F}^{cb}+E_F^{cb}}{\sqrt{M_{cb}^{*2}+2\nu|q_{cb}|B}}\Bigg), \nonumber \\
\rho_s^{ub}&=\frac{M_{ub}^{*2}}{2\pi^2}\Bigg[E_F^{ub} k_F^{ub} -  M_{ub}^{*2} \ln\Bigg(\frac{k_{F}^{ub}+E_F^{ub}}{M_{ub}}\Bigg)\Bigg]\ ,  \nonumber \\
\rho_{cb}&= \frac{|q_{cb}|B}{2\pi^2}\sum_{\nu=0}^{\nu_{max}}r_{\nu} k_{F}^{cb}, \nonumber \\
\rho_{ub}&=\frac{(k_F^{ub})^3}{3\pi^2},\nonumber \\
\rho_l &=\frac{|q_l|B}{2\pi^2}\sum_{\nu=0}^{\nu_{max}}r_{\nu} k_{F}^l\ ,
\end{align}
where $r_{\nu}$ is the Landau degeneracy of $\nu$ level. For  hadronic matter, the spin degeneracy is 2 for all Landau levels, except for the ground state, $\nu=0$, in which case it is 1. $k_{F}^{cb,ub,l}$ represents the Fermi momentum of charged baryons, uncharged baryons, and leptons, respectively. 
The pressure in the presence of a magnetic field is obtained from the energy density as
\begin{equation}
P_m = \sum_{i} \mu_i \rho_i^v -\mathcal{E}_m = \mu_n \sum_{b} \rho_b^v -\mathcal{E}_m \ , 
\end{equation}
where the last equality is obtained using charge neutrality and $\beta$-equilibrium conditions.
For quark matter, the presence of the magnetic field in the energy density and pressure modifies the first term of Eq. (\ref{q1}) as
\vspace{-0.3cm}
\begin{align}
\mathcal{E}_f&= \sum_{i=u,d,s,e}\Biggl\{\frac{|q_{i}|B}{4\pi^2}\sum_{\nu=0}^{\nu^i_{max}}g_{i}  \times  \Bigg[\mu_i \sqrt{\mu_i^2 -(M_{\nu}^i)^2}-(M_{\nu}^i)^2 \ln\Bigg(\frac{\mu_i + \sqrt{\mu_i^2 +(M_{\nu}^i)^2}}{(M_{\nu}^i)^2}\Bigg)\Bigg]\Biggr\}\ , \nonumber \\
P_f&= \sum_{i=u,d,s,e}\Biggl\{\frac{|q_{i}|B}{4\pi^2}\sum_{\nu=0}^{\nu^i_{max}}g_{i}  \times \Bigg[\mu_i \sqrt{\mu_i^2 -(M_{\nu}^i)^2}-(M_{\nu}^i)^2 \ln\Bigg(\frac{\mu_i + \sqrt{\mu_i^2 -(M_{\nu}^i)^2}}{(M_{\nu}^i)^2}\Bigg)\Bigg]\Biggr\}\ .
\end{align}
The quark density in presence of a strong magnetic field is
\begin{align}
\rho_q&= \sum_{i=u,d,s}\Biggl\{\frac{|q_{i}|B}{2\pi^2}\sum_{\nu=0}^{\nu^i_{max}}g_{i}  \sqrt{\mu_i^2 -(M_{\nu}^i)^2} \Biggr\}\ ,
\end{align}
where $M_{\nu}^i = \sqrt{m_i^2 + 2\nu |q_i|B}$. The factor $g_i$ represents the degeneracy of $i$th particle, which is 6 for quarks and 2 for leptons, except for the zeroth Landau level, in which case it is 3 and 1, respectively. The baryon density $\rho_b$ is defined as $\rho_b = \sum_q \rho_q /3$ and 
\begin{align}
\rho_3 &= \sum_{q}\psi^{\dagger}\tau_3\psi =\rho_u -\rho_d \ .
\end{align}
\bibliography{ref}{}
\bibliographystyle{aasjournal}
\end{document}